\documentclass[lettersize,journal]{IEEEtran}

\usepackage{amsmath,amsfonts}
\usepackage{algorithmic}
\usepackage{graphicx}
\usepackage{textcomp}
\usepackage[dvipsnames,table]{xcolor}
\usepackage{xspace}
\usepackage{listings}
\usepackage[inline]{enumitem}
\usepackage{booktabs}
\usepackage{makecell}
\usepackage{tcolorbox}
\usepackage{balance}
\usepackage{amsthm}
\usepackage{tikz}
\usepackage{multirow}
\usepackage{pifont}% http://ctan.org/pkg/pifont
\usepackage{subcaption}
\usepackage{utfsym}
\usepackage{tcolorbox}
\tcbuselibrary{listings, breakable, skins}
\usepackage{hhline}
\usepackage{colortbl}
\usepackage{array}
\usepackage{pifont}
\usepackage{hyperref}

\definecolor{lavender}{rgb}{0.9, 0.9, 0.98}
\newcommand*\np[2][z]{%\textcolor{red}{%
\ifx z#1%
$\numprint{#2}$%
\else%
$\numprint[#1]{#2}$%
\fi\xspace%}
}

\begin{document}

\renewcommand{\texttt}[1]{{\small\ttfamily#1}}
\renewcommand{\path}[1]{{\small\ttfamily#1}}

\AddToHook{env/tabular/begin}{
    \renewcommand{\texttt}[1]{{\ttfamily #1}}
    \renewcommand{\path}[1]{{\ttfamily #1}}
}

\newcommand{\manifest}{dependency specification file\xspace}
\newcommand{\etal}{\textit{et al.}\xspace}

\newcommand{\toolName}{\textsc{fika}\xspace}
\newcommand{\semgrep}{Semgrep\xspace}
\newcommand{\fuzzer}{Jazzer\xspace}

\newcommand{\symboltpl}{$m_{tpl}$\xspace}
\newcommand{\symboldp}{$m_{dp}$\xspace}
\newcommand{\symbole}{$m_e$\xspace}
\newcommand{\symbolt}{$m_t$\xspace}
\newcommand{\symbold}{$m_d$\xspace}

\newcommand{\todo}[1]{\textcolor{red}{#1}}
\newcommand{\todoyogya}[1]{\textcolor{pink}{#1}}
\newcommand{\todomeriem}[1]{\textcolor{Emerald}{#1}}
\newcommand{\modified}[1]{{\color{black}#1}}

\newcommand{\dependency}{third-party library\xspace}
\newcommand{\Dependency}{Third-party library\xspace}
\newcommand{\DEpendency}{Third-party Library\xspace}
\newcommand{\dependencies}{third-party libraries\xspace}
\newcommand{\Dependencies}{Third-party libraries\xspace}
\newcommand{\model}{DeepSeek V3.2\xspace}
\newcommand{\snippets}{source code snippets\xspace}
\newcommand{\snippet}{source code snippet\xspace}
\newcommand{\nbiterations}{5\xspace}
\newcommand{\nbprojects}{8\xspace}
\newcommand{\harness}{reachability scenario\xspace}
\newcommand{\harnesses}{reachability scenarios\xspace}
\newcommand{\Harness}{Reachability scenario\xspace}
\newcommand{\Harnesses}{Reachability scenarios\xspace}
\newcommand{\testGenEngine}{\harness generation pipeline\xspace}
\newcommand{\harnessMultiLine}{R.scenarios}
\newcommand{\executability}{executability\xspace}
\newcommand{\Executability}{Executability\xspace}
\newcommand{\executable}{executable\xspace}

\newcommand{\nbTotalNewCovTplCallSites}{609\xspace}
\newcommand{\nbTotalUncovTplCallSites}{1465\xspace}

\newcommand{\nbTotalCovTplCallSites}{2363\xspace}
\newcommand{\improvement}{20\%\xspace}
\newcommand{\testcov}{54\%\xspace}
\newcommand{\perHarnessCost}{0.00025\xspace}

%%%%%%%%%%%%%%%%%%%%%%%%%%%%%%%
%%%%%%%%%%%%%%%%%%%%%%%%%%%%%%%
%%%%%%%%%%%%%%%%%%%%%%%%%%%%%%%
%%% stats %%%

% Flink
\newcommand{\nbModuleFlink}{core}
\newcommand{\nbCommitFlink}{\href{https://github.com/apache/flink/tree/61f9ffece974134c8d7da8f87e3092b54af9a700}{61f9ffe}}
\newcommand{\nbLocFlink}{120k}
\newcommand{\nbEPFlink}{7463}
\newcommand{\nbTestsFlink}{7655}
\newcommand{\nbCoverageFlink}{66\%}
\newcommand{\nbDepsFlink}{8 (12)}

%  GraphHopper 
\newcommand{\nbModuleGraphhopper}{core}
\newcommand{\nbCommitGraphhopper}{\href{https://github.com/graphhopper/graphhopper/tree/1c811e5661af2a6d79ce900d620ee468558bda30}{1c811e5}}
\newcommand{\nbLocGraphhopper}{70k}
\newcommand{\nbEPGraphhopper}{3332}
\newcommand{\nbTestsGraphhopper}{2561}
\newcommand{\nbCoverageGraphhopper}{84\%}
\newcommand{\nbDepsGraphhopper}{10 (18)}

%  Jooby 
\newcommand{\nbModuleJooby}{jooby}
\newcommand{\nbCommitJooby}{\href{https://github.com/jooby-project/jooby/tree/d2272e7bba0e9c6c6f97827eccd2550c56d3a1cf}{d2272e7}}
\newcommand{\nbLocJooby}{25k}
\newcommand{\nbEPJooby}{2677}
\newcommand{\nbTestsJooby}{123}
\newcommand{\nbCoverageJooby}{24\%}
\newcommand{\nbDepsJooby}{5 (6)}

%  MyBatis 
\newcommand{\nbModuleMyBatis}{--}
\newcommand{\nbCommitMyBatis}{\href{https://github.com/mybatis/mybatis-3/tree/57c7c415f4eac68a996e0a9f2996b6b1ce426d97}{57c7c41}}
\newcommand{\nbLocMyBatis}{71k}
\newcommand{\nbEPMyBatis}{6139}
\newcommand{\nbTestsMyBatis}{1990}
\newcommand{\nbCoverageMyBatis}{87\%}
\newcommand{\nbDepsMyBatis}{8 (9)}

%  PDFBox 
\newcommand{\nbModulePdfBox}{pdfbox}
\newcommand{\nbCommitPdfBox}{\href{https://github.com/apache/pdfbox/tree/32600224305ee31a5a7859890a52e19c82458fa8}{3260022}}
\newcommand{\nbLocPdfBox}{102k}
\newcommand{\nbEPPdfBox}{5818}
\newcommand{\nbTestsPdfBox}{700}
\newcommand{\nbCoveragePdfBox}{60\%}
\newcommand{\nbDepsPdfBox}{3 (4)}

%  Tablesaw 
\newcommand{\nbModuleTablesaw}{json}
\newcommand{\nbCommitTablesaw}{\href{https://github.com/jtablesaw/tablesaw/tree/faf0d54d9ddab6ad5a04e20f8ee4d255c4ae5148}{faf0d54}}
\newcommand{\nbLocTablesaw}{879}
\newcommand{\nbEPTablesaw}{120}
\newcommand{\nbTestsTablesaw}{14}
\newcommand{\nbCoverageTablesaw}{70\%}
\newcommand{\nbDepsTablesaw}{3 (22)}

%  Tika 
\newcommand{\nbModuleTika}{core}
\newcommand{\nbCommitTika}{\href{https://github.com/apache/tika/tree/bb785a220c9861db5b279dd8a3835004963a695e}{bb785a2}}
\newcommand{\nbLocTika}{30k}
\newcommand{\nbEPTika}{1629}
\newcommand{\nbTestsTika}{309}
\newcommand{\nbCoverageTika}{46\%}
\newcommand{\nbDepsTika}{5 (16)}

%  Poi-tl 
\newcommand{\nbModulePoiTl}{poi-tl}
\newcommand{\nbCommitPoiTl}{\href{https://github.com/Sayi/poi-tl/tree/58fdb6c7d9db7da53cd420ff1ebde49813bd7af2}{58fdb6c}}
\newcommand{\nbLocPoiTl}{30k}
\newcommand{\nbEPPoiTl}{1488}
\newcommand{\nbTestsPoiTl}{127}
\newcommand{\nbCoveragePoiTl}{78\%}
\newcommand{\nbDepsPoiTl}{5 (36)}

%%%%%%%%%%%%%%%%%%%%%%%%%%%%%%%
%%%%%%%%%%%%%%%%%%%%%%%%%%%%%%%
%%%%%%%%%%%%%%%%%%%%%%%%%%%%%%%
\newcommand{\projFlink}{flink}
\newcommand{\projGraphhopper}{graphhopper}
\newcommand{\projJooby}{jooby}
\newcommand{\projMybatis}{mybatis-3}
\newcommand{\projPdfbox}{pdfbox}
\newcommand{\projTablesaw}{tablesaw}
\newcommand{\projTika}{tika}
\newcommand{\projPoiTl}{poi-tl}

\newcommand{\nbUniqueTPLMethods}[1]{#1}
\newcommand{\nbAllCallSites}[1]{#1}
\newcommand{\nbReachableCallSites}[1]{#1}
\newcommand{\nbReachableCallSitesPerc}[1]{#1} % 

\newcommand{\nbCoveredByExistingTests}[1]{#1}
\newcommand{\nbAllCoveredByExistingTests}[1]{#1} % 

\newcommand{\nbCoveredByNewTests}[1]{#1}
\newcommand{\nbOnlyCoveredDynamicallyTotal}[1]{#1}
\newcommand{\nbCoveredByTotal}[1]{#1}
\newcommand{\nbIncrease}[1]{#1}

\newcommand{\nbCoveredDynamically}[1]{#1}

\newcommand{\nbTestGenAttempts}[1]{#1}
\newcommand{\nbSuccessfulAttempts}[1]{#1}
\newcommand{\nbWeakest}[1]{#1}
\newcommand{\nbLessWeak}[1]{#1}
\newcommand{\nbRandom}[1]{#1}
\newcommand{\nbFikaPrompt}[1]{#1}
\newcommand{\nbIun}[1]{#1}
\newcommand{\nbIdeux}[1]{#1}
\newcommand{\nbItrois}[1]{#1}
\newcommand{\nbIquatre}[1]{#1}
\newcommand{\nbIcinq}[1]{#1}

\newcommand{\nbCve}[1]{#1}
\newcommand{\nbVultpl}[1]{#1}
\newcommand{\nbRsemgrep}[1]{#1}
\newcommand{\nbDir}[1]{#1}
\newcommand{\nbEfika}[1]{#1}
\newcommand{\nbUsemgrep}[1]{#1}
\newcommand{\nbSRsemgrep}[1]{#1}
\newcommand{\nbLRsemgrep}[1]{#1}

\newcommand{\cve}[1]{#1}
\newcommand{\fmodule}[1]{#1}
\newcommand{\ftpl}[1]{#1}
\newcommand{\reachsemgrep}[1]{#1}
\newcommand{\reachfika}[1]{#1}
\newcommand{\undetsemgrep}[1]{#1}
\newcommand{\reachstrongsemgrep}[1]{#1}
\newcommand{\reachweaksemgrep}[1]{#1}
\newcommand{\reachstaticfika}[1]{#1}

\newcommand{\nbstatic}[1]{#1}
\newcommand{\nbtest}[1]{#1}
\newcommand{\nbgen}[1]{#1}
\newcommand{\nbexec}[1]{#1}
\newcommand{\nbapi}[1]{#1}

%%%%%%%%%%%%%%%%%%%%%%%%%%%%%%%
%%%%%%%%%%%%%%%%%%%%%%%%%%%%%%%
%%%%%%%%%%%%%%%%%%%%%%%%%%%%%%%

\definecolor{shade1}{RGB}{255,220,150} % yellow (for SR)
\definecolor{shade2}{RGB}{255,190,120} % light orange (for LR)
\definecolor{shade3}{RGB}{255,160,70}  % brighter orange (for U)
\definecolor{shade4}{RGB}{255,240,70} % yellow again (for fika)

% Source - https://tex.stackexchange.com/a/488962
% Posted by Steven B. Segletes, modified by community. See post 'Timeline' for change history
% Retrieved 2026-04-09, License - CC BY-SA 4.0
\newcommand*\emptycirc[1][0.7ex]{\tikz[baseline=-0.5ex]\draw(0,0) circle (#1);} 
\newcommand*\halfcirc[1][0.7ex]{%
  \tikz[baseline=-0.5ex]{
    \draw[fill] (0,0)-- (90:#1) arc (90:270:#1) -- cycle;
    \draw (0,0) circle (#1);
  }}
\newcommand*\fullcirc[1][0.7ex]{\tikz[baseline=-0.5ex]\fill (0,0) circle (#1);} 
\DeclareRobustCommand{\newcheckmark}{%
  \tikz[baseline=-0.5ex]{%
    \fill[scale=0.27]
      (0,.35) -- (.25,0) -- (1,.7) -- (.25,.15) -- cycle;
  }%
}
\DeclareRobustCommand{\boldcheckmark}{%
  \tikz[baseline=-0.5ex]{%
    \fill[scale=0.30]
      (0,.35) -- (.25,0) -- (1,.77) -- (.25,.27) -- cycle;
  }%
}
\lstdefinestyle{customjava}{
    language=Java,
    basicstyle=\ttfamily\scriptsize,
    keywordstyle=\color{blue},
    commentstyle=\color{gray},
    stringstyle=\color{orange},
    showstringspaces=false,
    tabsize=4,
    breaklines=true,
    numbers=left,
    numberstyle=\tiny\color{gray},
    xleftmargin=2em,
    frame=single
}
\lstdefinelanguage{json}{
    basicstyle=\ttfamily\small,
    showstringspaces=false,
    breaklines=true,
    frame=single,
    morestring=[b]",
    morecomment=[l]{//},
    morekeywords={:,{}},
    keywordstyle=\color{black},
    stringstyle=\color{blue}
}
\lstdefinelanguage{json}{
    basicstyle=\ttfamily\small,
    showstringspaces=false,
    breaklines=true,
    frame=single,
    morestring=[b]",
    morecomment=[l]{//},
    stringstyle=\color{blue},       % JSON values
    keywordstyle=\color{black},     % punctuation
    escapeinside={(*@}{@*)}         % escape to LaTeX for coloring keys
}

\lstdefinelanguage{jsondiff}{
  basicstyle=\ttfamily\small,
  breaklines=true,
  breakatwhitespace=false,
  showstringspaces=false,
  escapeinside={(*@}{@*)}, % Allows manual formatting
  numbers=left,
  numberstyle=\tiny\color{gray},
  xleftmargin=2em,
  frame=single
}

\lstdefinestyle{customjson}{
    language=json,
    basicstyle=\ttfamily\small,
    showstringspaces=false,
    frame=none,
    escapechar=!,
    literate={
        {\"}{{\textcolor{blue}{\texttt{"}}}}1
        {:}{{\textcolor{red}{\texttt{:}}}}1
        {,}{{\textcolor{gray}{\texttt{,}}}}1
        {[}{{\textcolor{purple}{\texttt{[}}}}1
        {]}{{\textcolor{purple}{\texttt{]}}}}1
        {\{}{{\textbf{\texttt{\{}}}}1
        {\}}{{\textbf{\texttt{\}}}}}1
    }
}

% Numbered, autoref-able tcolorbox
\newtcolorbox[auto counter, number within=section]{promptbox}[2][]{
    % breakable,
    enhanced,
    colback=white,       % background of the box
    colframe=black,      % frame color
    colbacktitle=white,  % background of the title
    coltitle=black,      % title text color
    boxrule=0.8pt,
    sharp corners,
    title={#2},
    label={#1},
    fonttitle=\bfseries,
    left=2mm,
    right=2mm,
    top=0mm,
    bottom=0mm,
}

\definecolor{keywordcolor}{RGB}{0,102,204}
\definecolor{stringcolor}{RGB}{196,26,22}
\definecolor{commentcolor}{RGB}{0,128,0}
\definecolor{bgcolor}{RGB}{250,250,250}
\cellcolor[rgb]{0.961,0.957,0.996}
\definecolor{rowGray}{rgb}{0.91,0.918,0.918}

\lstdefinestyle{algorithmstyle}{
    backgroundcolor=\color{bgcolor},
    basicstyle=\ttfamily\small,
    keywordstyle=\color{keywordcolor}\bfseries,
    commentstyle=\color{commentcolor}\itshape,
    stringstyle=\color{stringcolor},
    showstringspaces=false,
    breaklines=true,
    captionpos=b,
    frame=single,
    rulecolor=\color{gray},
    frameround=tttt,
    aboveskip=10pt,
    belowskip=10pt
}

% git diffs
% \lstset{
%   literate={+}{{{\color{green}+}}}1
%            {-}{{{\color{red}-}}}1
% }

\title{\toolName: Expanding Dependency Reachability with Executability Guarantees}

\author{
    \IEEEauthorblockN{
        Yogya Gamage\IEEEauthorrefmark{1},
        Meriem Ben Chaaben\IEEEauthorrefmark{1},
        Martin Monperrus\IEEEauthorrefmark{2},
        Benoit Baudry\IEEEauthorrefmark{1}
    } \\
    \IEEEauthorblockA{\IEEEauthorrefmark{1}Université de Montréal, Montréal, Canada} \\
    \IEEEauthorblockA{\IEEEauthorrefmark{2}KTH Royal Institute of Technology, Stockholm, Sweden} \\
    \IEEEauthorrefmark{1}{\{yogya.gamage,meriem.ben.chaaben,benoit.baudry\}@umontreal.ca}
    \IEEEauthorrefmark{2}{monperrus@kth.se} 
    %\\Email: \{marcin.copik, alexandru.calotoiu, htor\}@inf.ethz.ch}
    %\\Email: kotaranov@microsoft.com}
}

% Tracing Third-Party API Reachability with LLMs
\maketitle

\begin{abstract}
Automated \dependency analysis tools help developers by addressing key dependency management challenges, such as automating version updates, detecting vulnerabilities, and detecting breaking updates. 
Dependency reachability analysis aims at improving the  precision of dependency management, by reducing the space of dependency issues to the ones that actually matter.
Most tools for dependency reachability analysis are static and fundamentally limited by the absence of execution.
In this paper, we propose \toolName, a pipeline for providing guarantees of \executability for \dependency call sites. \toolName leverages a Large Language Model (LLM) guided by static analysis to generate code that is executed, and whose execution trace provides guarantees that a \dependency call site is actually \executable. 
We apply our approach to a dataset of eight Java projects to empirically evaluate the effectiveness of \toolName. On average, \testcov of these call sites are covered by the existing test suites, and therefore, have evidence for their \executability. \toolName improves this coverage by \improvement, generating successful \harnesses for \nbTotalNewCovTplCallSites out of \nbTotalUncovTplCallSites previously non-covered call sites.
In six out of eight projects, \toolName provides strong guarantees that more than 75\% of call sites are \executable.
\modified{We further demonstrate that \toolName improves over \semgrep, a state-of-the-art static vulnerability reachability analysis tool, and outperforms \fuzzer, a coverage-guided directed fuzzer.} We show that \toolName can help prioritize the vulnerability updates with stronger guarantees of \executability.
\end{abstract}

\section{Introduction}
When developing software, developers write code and import \dependencies, that are fetched from package registries through package managers \cite{Cox19}. 
%This code is then built into reusable libraries or executable packages by build systems. 
Software reuse is essential for modern software development \cite{mohagheghi2007quality}, and yet, when \dependencies are not properly managed, it can lead to a range of software maintenance challenges. These include dependency incompatibilities \cite{rausch2017empirical}, security vulnerabilities \cite{ladisa2023sok}, breaking updates \cite{frank2024bump}, and phantom dependencies that are used in the code but not properly declared in dependency configuration files \cite{larson2025phantom}. These issues can ultimately result in build failures, production bugs, or catastrophic software supply chain attacks such as Log4Shell \cite{log4shell}.

%manifest only
% api only - only be as accurate as the underlying the analysis technique
% method signatures for breaking updates fro compilation
% different levels of granulariy
% from manifest to whole code
% if you get a precise analysis of the interactions then you get a btter analsis
In the past decade, several methods and tools have emerged to address the challenges of dependency management \cite{williams2025dir}. 
Automated version update tools notify developers when their \dependencies are outdated \cite{Mohayeji2025dependabot}, and vulnerability scanning tools alert developers when a known vulnerable version is present among the dependencies of a project \cite{henrik2022vul}. 
These tools rely on  lockfiles \cite{gamage2026design} or Software Bills of Materials (SBOMs) \cite{balliu2023challenges} to determine the exact list of \dependencies for a project.
As their notifications are based solely on the names and versions of the dependencies rather than on the actual usage of libraries by the project codebase, they often produce false positives, and create notification fatigue for developers\unskip ~\cite{mirhosseini2017fatigue}.

In order to focus dependency management on the actual interactions between a project and its dependencies, the recent tools have introduced a step of \dependency reachability analysis. This step aims to identify the \dependency methods that are actually used. State-of-the-art dependency reachability analysis relies on the static analysis of complete dependency codebases \cite{cesar2021debloating, keshani2024frankenstein}. While this increases the precision of dependency management, it can still produce false alarms \cite{sui2018soundness}, since static reachability does not guarantee that a third-party method is truly executable from the project's codebase.

\begin{figure*}[t]
    \centering
    \includegraphics[width=0.9\linewidth]{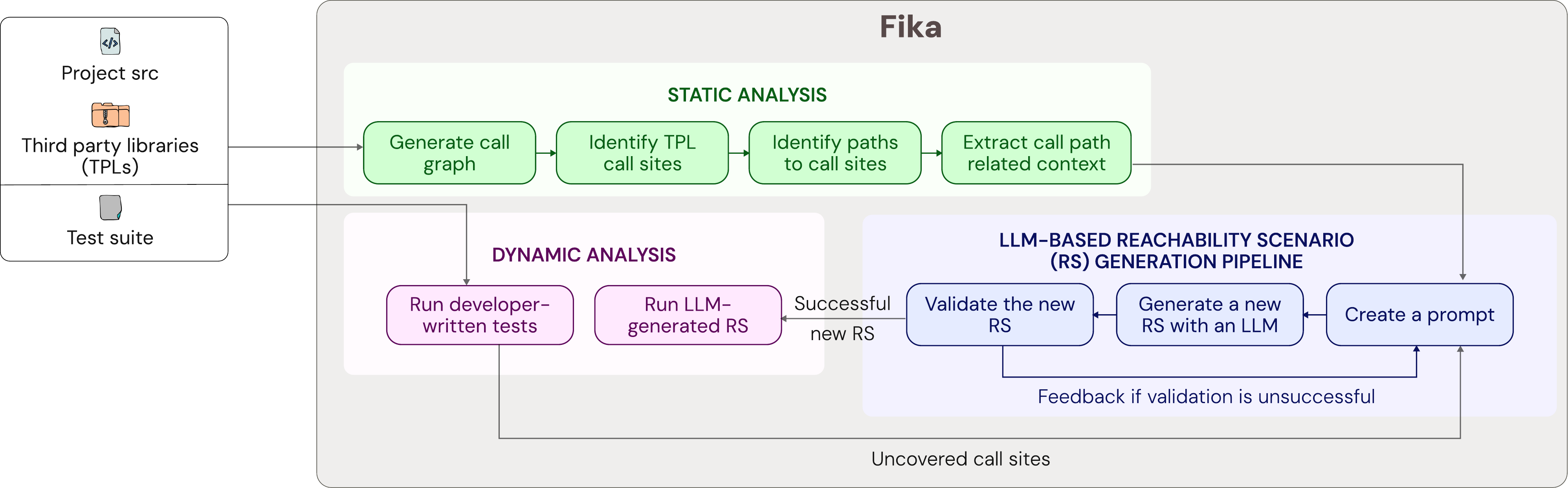}
    \captionsetup{font=small}
    \caption{The architecture of \toolName. It takes the project source code as input. Static analysis produces a set of call sites of \dependencies, which is then used to create \harnesses for invoking non-covered call sites. The generated \harnesses are validated and integrated following successful validation.}
    \label{fig:diagram-both}
\end{figure*}

In this paper, we propose an end-to-end automated pipeline, called \toolName, to improve the guarantees that a given \dependency is truly executable in an application.
In case of success, \toolName produces a concrete piece of code which we call a \textit{\harness} that  can be executed from the client project to demonstrate \executability. 
\toolName's first step performs static analysis of the client project's code to identify all invocations of third-party methods, i.e. \dependency call sites. 
Then, \toolName checks whether a static path exists from a public method to such a \dependency call site. 
The second step executes the  test suite of the project to determine which of the \dependency call sites are already \executable, hence need no further care. 
The third step aims at increasing the guarantees of \executability of \dependency call sites. For each call site that is not reached by an existing test, we generate a new \harness with a Large Language Model (LLM) based pipeline. We prompt the LLM to generate a \harness for the public method that eventually triggers the \dependency call site, with the sole goal of covering that target call site. We leverage the static code analysis and dynamic analysis from the two first steps to provide rich context for the LLM to generate a \harness.
To demonstrate the feasibility of our end-to-end pipeline, we implement the approach for Java projects.

% evaluation
We evaluate the effectiveness of our tool by analyzing 8 client projects and their \dependencies. Our results show that, on average, \testcov of third-party method invocations are covered by the existing test suite. We demonstrate that \toolName can generate successful \harnesses for \nbTotalNewCovTplCallSites out of \nbTotalUncovTplCallSites previously non-covered \dependency call sites. Overall, \toolName increases the \executability guarantees by, on average, \improvement.

To evaluate \toolName in a real-world setting, we analyze 13 modules with reported vulnerabilities. \modified{We compare \toolName against \semgrep, a state-of-the-art vulnerability reachability analysis tool, and \fuzzer, a coverage-guided directed fuzzer. \toolName confirms the \executability of vulnerable \dependencies in cases where \semgrep produces ambiguous results, and outperforms \fuzzer, thereby helping developers prioritize vulnerability fixes.}
% novelty claim
To the best of our knowledge, \toolName is the first tool capable of improving the \executability analysis of \dependencies through automated \harness generation.
% State-of-the-art reachability analysis for third-party dependencies is only based on static analysis. 

In summary, our key contributions are as follows.
\begin{itemize}[leftmargin=*]
    \item An architecture for generating \harness that provably reach dependencies and trigger \dependency interactions. 
    \item A publicly available prototype tool \toolName for Java to expand the evidence of \executability of Maven \dependency call sites.
    \item An empirical evaluation over eight Java projects, demonstrating that \toolName can reach \nbTotalNewCovTplCallSites out of \nbTotalUncovTplCallSites previously non-covered dependency call sites.
    \item \modified{A demonstration that  \toolName's \harness generation clearly improves the precision of vulnerable dependency reachability compared to the state-of-the-art reachability analysis tool \semgrep and \fuzzer, a coverage-guided directed fuzzer.}
\end{itemize}

\section{\toolName}
 
This section describes \toolName, a framework that extends the guarantees of \executability for third-party library call sites within a project. These guarantees enable the explicit identification of \dependency usage.

\subsection{Key concepts}

We first introduce terminology we use throughout the paper, and present the related concepts to the design of \toolName.

\textbf{Project}
A project ($p$) is a collection of source code files written in an object oriented programming language. The project declares dependencies on external libraries or frameworks in dependency specification file (e.g. pom.xml in Maven).

\textbf{Entry Point}
An entry point (\symbole) is a publicly accessible method of a project that allows the code to be invoked by other projects or by a main function. These public methods are generally the targets for writing test cases, as they represent the intended usage of the project’s functionality.

\textbf{\Dependency}
A \dependency ($tpl$) is a reusable set of functions developed by external parties, available on a package registry. A project ($p$) depends on a \dependency ($tpl$) either because it lists $tpl$ in its dependency specification file (direct dependency), or a \dependency may be resolved indirectly for the project via other declared \dependencies (indirect dependency). 

\textbf{Reachability of a \dependency method}
\modified{A \dependency method is reachable from a client method if (i) there exists a \emph{call path} from a given entry point method (\symbole) in the client project ($p$) to a target method (\symbolt) in a \dependency ($tpl$); and (ii) the call path is executable, i.e., there exists at least one execution of the client that covers this path from \symbole to \symbolt.}

\textbf{Call Path}
In the context of reachability analysis, a call path consists of the entry point \symbole (root of the path), the target method \symbolt (end of the path), and additional methods between \symbolt and \symbole. 
We denote the direct caller of the target \symbold.
If the entry point directly calls the target, then \symbole and \symbold refer to the same method.  

\textbf{\Dependency call site}
We define a \dependency call site as any direct invocation to a method available in its \dependencies, within the project ($p$). Such a call site is located in  method \symbold of ($p$), and invoked method  \symbolt is defined in \dependency ($tpl$). Therefore, we denote them as \symboldp and \symboltpl, respectively.
One \dependency method \symboltpl, may have multiple call sites within a project $p$, as there can be multiple \symboldp that call the same \symboltpl.

Because all libraries added to the project classpath can be used by the project, we do not distinguish between methods coming from direct \dependencies and those added indirectly. If any method from a direct or indirect \dependency is invoked  from the project, we consider this as a direct \dependency call site. Indirect call sites, where a \dependency method is invoked via another \dependency method, are out of scope for this study.

\textbf{\Harness}
We call \harness a code snippet that invokes a \symbole and initializes the project state in order to trigger a call path reaching the target \symboldp and invoke a \symboltpl . To execute a \harness in isolation, we implement, and run it within a unit testing framework, and use test coverage tools to confirm \executability. 
\modified{While \harnesses and unit tests are both executed with the same framework, their intent and usage are fundamentally different: test cases written by developers are intended to find bugs and hence include both a test input and a test oracle that define expectations about the program's behavior; \harnesses are intended to show that there exists an executable path that can reach a \dependency call site. They hence include a complete initialization state for this executable path and do not contain any assertion.}
% This resembles a unit test case, except that its purpose is to collect evidence of \dependency call site \executability, and hence a \harness does not include any assertion.

\subsection{Overview}

We present an overview of \toolName in \autoref{fig:diagram-both}. The system takes the project as input. In the static reachability analysis phase, \toolName  generates a static call graph for the project. It then generates a detailed record of statically reachable \dependency call sites and the associated context details for those call sites (see \autoref{sec:context_extraction}).  
In its dynamic analysis phase, \toolName runs the project's test suite and then outputs the list of \dependency call sites that are covered by the existing test suite and therefore have evidence of \executability.

Every call site that is not covered is passed as a target to an LLM-based \testGenEngine, whose goal is to generate a \harness for covering the target. In the validation step, \toolName builds and runs the generated \harnesses. 
In case of success, the new \harness covers a new target and provides proof of the \executability of the corresponding \dependency call site. After this step, \toolName produces an extended list of newly covered \dependency call sites with evidence of \executability.
\Executability analysis is important for the precise analysis of \dependency interactions. 
In the following subsections, we describe the steps of \toolName in detail.

\subsection{Static analysis}

Given a  project ($p$) and its dependency specification file, \toolName's  static analysis phase aims at identifying all  \dependency call sites, as well as the candidate paths from the public methods (\symbole)  of $p$ that invoke these methods.

\subsubsection{Call graph generation}
 
We generate a static call graph for the project by considering all public methods within the project as entry points. The call graph is dependency wide, it includes all dependency classes in the call graph.
There are three main static analysis algorithms for creating call graphs in object oriented languages: Class Hierarchy Analysis (CHA), Rapid Type Analysis (RTA), and Variable Type Analysis (VTA) \cite{sui2018soundness}.
CHA identifies potential target methods based only on the declared types of variables and the class hierarchy. RTA applies more filters and limits potential target methods to classes that are more likely to be instantiated. VTA is a further specialized algorithm that tracks the flow of variables and considers pointer behavior.
We use the CHA algorithm to construct the call graph because, compared to RTA and VTA, CHA does not remove edges based on the likelihood of instantiation and therefore misses the fewest edges in the resulting call graph \cite{samhi2024dynamic}. 

\subsubsection{Identify paths to call sites}

In this step, \toolName computes the call paths that reach \dependency call sites, consisting of a pair of \symboltpl and \symboldp, from public methods. We first reverse the generated call graph to represent the callee-to-caller relationship. Next, we look for \dependency call sites within the project. In any call site, if the \symboldp is not a public method, we traverse the call graph breadth-first until we find a public method, which we consider as a potential entry point (\symbole). This breadth-first-search gives us one of the shortest paths to reach a public method from a non-public \symboldp.
% We record all shortest call paths that reach a \dependency call site from a public method of the project under study (\symbole). 
Note that there might be multiple paths from one public method to a call site. However, to handle the potential explosion in the number of paths, for further dynamic analysis, we select only one static path per tuple (\symbole, \symboldp, \symboltpl): the first path through a breadth-first search, and therefore a shortest path.

\subsubsection{Call path context extraction}
\label{sec:context_extraction}

Once \toolName has detected call paths, we extract additional context details about each path. This context is used in the subsequent steps of \toolName related to dynamic analysis and \harness generation with LLMs.

First, we extract the source code for methods along the call path as \snippets. Then, we annotate the \snippets by adding a comment before each line that includes a method invocation that is part of the call path.

%We also instrument the methods,  adding a comment before each line in order to track the path easily 
Second, we collect everything that is necessary to initialize the project entry point and invoke the public method (\symbole): the constructors, the instance or class variables of the class to which the entry-point method (\symbole) belongs. If the constructors are not public, we collect public factory methods. 
% Too late to collect all factory methods. We can justify this based on the fact that we also do not want to make the prompt longer due to the context window limits
We also collect methods, if any, that set values for instance or class variables.  
%We collect this context information about the entry point (\symbole) because these details are needed to initialize an object of the class in order to call the entry point method.

At the end of this step, the extracted context consists of a set of \snippets: code of methods along the call paths, code of the constructors and setters of \symbole, and all instance or class variables and imports related to the invocation of \symbole. This is a big chunk of textual information about the call path under consideration.

\subsection{Dynamic analysis}

The goal of \toolName's dynamic analysis is to determine which call paths identified during the static analysis are executable. 

We first check whether the statically identified call sites are already covered by the developer-written test suite of the project. To do so, we execute the existing tests and collect coverage details. We consider all the \dependency call sites that are part of the covered statements as executable.

For the non-covered call sites, we generate new \harnesses. These scenarios are automatically generated with a novel LLM-based \testGenEngine, as explained in the subsequent section. Once the generation is completed, we execute the successfully generated \harnesses, as a means to exercise non-covered call paths. The combination of developer-written tests and \harnesses provides concrete guarantees about the \executability of \dependency call sites in the project.

\subsection{LLM-based reachability scenario generation}

For each call site not covered by developer-written tests, we generate a \harness with the help of an LLM, with the goal of dynamically reaching the call site.

\subsubsection{\Harness generation}

We first create a structured prompt that provides the LLM with all information that we extracted in the static analysis phase: 
(i) the ordered call path from the public entry-point method (\symbole) to the target \dependency call site, 
(ii) the context (source code of all project methods along this path, instantiation information, see \autoref{sec:context_extraction}). 
The goal of LLM-based \testGenEngine is to generate a \harness that calls the entry-point (\symbole) with the goal of reaching the \dependency call sites. 
We instruct the LLM to not include any assertions, avoid modifying existing methods, and minimize mocking as mocks may create potentially infeasible call paths. 
The complete prompt is available on \href{https://github.com/sparkrew/fika/blob/main/pipeline/src/junit_agent/prompts.py}{GitHub}.

\subsubsection{\Harness validation}
\label{sec:validation}

In this step, \toolName validates the generated \harnesses against a set of criteria to verify their quality and ability to invoke the target \dependency call site. 

When a plausible \harness is generated, \toolName adds the \harness to the project and checks it against a set of static rules. 
\modified{This first set of rules aims at quickly eliminating generated \harnesses that  stub parts of the client project. We notice that LLMs can generate \harnesses that override some of the methods in the target path, or extend the class that we aim at executing. While these \harnesses will run correctly, they will not actually trigger the true behavior of the client project. This static filter eliminates \harnesses that match one of these three static patterns: they contain \textit{@Override}, they contain \textit{class ... extends ...}, or they contain  anonymous classes \textit{new ClassName() \{ ... \}}.}

If the static rules are satisfied, \toolName compiles and executes the generated \harness.
Recall that, to implement and execute a \harness, we rely on a testing framework, similar to how a unit test is executed within the project.
\toolName then verifies whether the \harness compiles successfully and whether the target \dependency call site is reached based on code coverage.

\subsubsection{\modified{Iterative feedback loop}}
\label{sec:feedback}

\modified{
\toolName uses an iterative feedback process in which the execution result of each generated \harness is used to guide the generation of the next \harness.

If any statically enforced rule is violated, \toolName generates targeted feedback describing the violation and integrates it into the next prompt for the \testGenEngine. For compilation errors or execution errors, \toolName extracts the relevant diagnostic output from the logs and includes the lines marked with error-level indicators (such as \textit{[ERROR]} in Maven) in the feedback. When a \harness executes successfully but does not reach the target call site, \toolName reports which methods along the call path are covered and where execution diverges.

To determine the iteration limit and how feedback from previous iterations should be retained, we conduct preliminary experiments on two projects, \texttt{Apache PDFBox} and \texttt{GraphHopper}. For each project, we run \toolName and generate \harnesses for five randomly selected paths. These experiments allow us to observe how many iterations are needed on average to reach the target call site and how retaining feedback from previous iterations affects the feedback process. 

We find that concatenating outputs from multiple iterations substantially increases the number of tokens passed to the \testGenEngine, growing with each additional iteration. For example, in PDFBox, one of the paths in our experiments reaches 32K tokens when the outputs and feedback from all iterations are concatenated, compared to only 5K tokens for the feedback of the latest iteration alone. This growth increases the computational cost and latency of each regeneration step, as the LLM processes more input tokens per call, with little corresponding benefit for the final \harness.
%since the target of each iteration is determined by the most recent execution outcome, earlier iterations' feedback becomes redundant once the \testGenEngine has already acted on it.
Moreover, prior work has shown that language models can struggle to make effective use of relevant information when it is embedded in a long context ~\cite{liu2024lost} and that their outputs can be degraded by the presence of irrelevant or outdated context~\cite{shi2023large}. 
Accumulating stale feedback from earlier iterations in the prompt can therefore dilute the \testGenEngine's attention and cause it to lose track of its actual goal of reaching the target method. Therefore, we provide only the feedback from the immediately preceding iteration rather than concatenating the outputs of all previous iterations.
 
Based on the same preliminary experiments, we set the maximum number of regeneration attempts to \nbiterations. In our pilot, we observe that additional iterations beyond this threshold did not result in successful \harness generation, and this does not justify the increased computational cost.
}

\subsection{Illustrative Example}
We present an end-to-end example in which we analyze the \textit{Graphhopper} project with \toolName, identify a non-covered \dependency call site, and generate a new \harness that can reach it. 

\textit{Graphhopper} is a popular open-source routing library and web server for OpenStreetMap data. Its core module provides routing functionality, and the class \textit{CHPreparationGraph}, within the \textit{routing} package, represents the graph data structure used for fast long-distance route calculations.

When analyzing \textit{Graphhopper}, \toolName determines that the \dependency method (\symboltpl), \path{IndirectSort.\allowbreak mergesort} (\autoref{fig:path-context-target}), from dependency \path{com.carrotsearch.\allowbreak hppc}, which is called by the direct caller (\symboldp) \path{ com.graphhopper.\allowbreak routing.\allowbreak ch.CHPreparationGraph.\allowbreak OrigGraph.\allowbreak Builder.\allowbreak build} (\autoref{fig:path-context-caller}), is not covered by any existing test. \toolName also identifies that this call site can be statically reached through the entry point (\symbole) \path{com.graphhopper.\allowbreak routing.ch.\allowbreak CHPreparationGraph.\allowbreak prepareForContraction} (\autoref{fig:path-context-entry}) in two hops.
In the call path context extraction step, \toolName extracts the \snippets of methods along the call path as shown in Figure \autoref{fig:example-case-methods}, and other \snippets that are required to invoke the \symbole as shown in \autoref{fig:example-case-extra}.
We include the content of both Figure \autoref{fig:example-case-methods} and \autoref{fig:example-case-extra} in the prompt.

The path to reach \symboltpl from \symbole contains one direct caller (\symboldp) in between. 
By prompting the LLM with the extracted context details, \toolName generates a \harness, shown in \autoref{fig:example-test}, that invokes \symbole and reaches \symboldp. 
This \harness creates an object of the entry point class \textit{CHPreparationGraph} by calling the factory method \textit{edgeBased}. The \textit{edgeBased} method requires three parameters: two integers and one object of type \textit{TurnCostFunction}. 
\toolName manages to create a real object, \textit{turnCostFunction} of the \textit{TurnCostFunction} class, and passes it along with the integers 5 and 10 to the \textit{edgeBased} method. \textit{edgeBased} then returns a graph object, which allows the \harness to invoke the entry point method \symbole, \textit{prepareForContraction()}. 

In the final step, static validation, compilation, and the target reachability check all succeed.
Consequently, this new generated \harness is a proof of \executability of \symboltpl,  \path{com.carrotsearch.\allowbreak hppc.sorting.\allowbreak IndirectSort.\allowbreak mergesort}, from the \textit{Graphhopper} project.
This confirmation can be used by \dependency analysis tools that rely on coverage, such as coverage-based debloating tools \cite{cesar2023coverage}, vulnerability analysis tools \cite{chen2024vultestlib}, and \dependency privilege detection tools \cite{amuso2025ztd}.

\begin{figure}[t]
\centering
\captionsetup{font=small}
\caption{An illustrative example where \toolName identifies an entry point method \symbole (highlighted in \colorbox{yellow}{yellow}) that statically invokes a method \symboldp (highlighted in \colorbox{pink}{pink}) that directly calls a \symboltpl (highlighted in \colorbox{green}{green}).
The example is taken from the class \href{https://github.com/graphhopper/graphhopper/blob/1c811e5661af2a6d79ce900d620ee468558bda30/core/src/main/java/com/graphhopper/routing/ch/CHPreparationGraph.java}{\texttt{CHPreparationGraph}} of the \textit{Graphhopper} project. All \snippets are extracted by \toolName, and are included in the prompt, except for the source code of \symboltpl, which is not part of the \textit{Graphhopper} project.}
\begin{subfigure}[t]{\linewidth}
\captionsetup{font=small}
\caption{\symbole\ From class CHPreparationGraph}
\label{fig:path-context-entry}
\begin{lstlisting}[style=customjava, escapechar=!, linewidth=\linewidth]
public void !\colorbox{yellow}{prepareForContraction()}! {
    checkNotReady();
    // PATH: Test should invoke the next CHPreparationGraph$OrigGraph$Builder.build(...) [step in execution path]
    origGraph = (edgeBased) ? origGraphBuilder.!\colorbox{pink}{build()}! : null;
    origGraphBuilder = null;
    ready = true;
}
\end{lstlisting}
\end{subfigure}

\begin{subfigure}[t]{\linewidth}
\captionsetup{font=small}
\caption{\symboldp\ From class CHPreparationGraph.OrigGraph.Builder}
\label{fig:path-context-caller}
\begin{lstlisting}[style=customjava, escapechar=!, linewidth=\linewidth]
OrigGraph !\colorbox{pink}{build()}! {
    // PATH: Test should invoke the next IndirectSort.mergesort(...) [step in execution path]
    int[] sortOrder =
        IndirectSort.!\colorbox{green}{mergesort}!(
        0, fromNodes.elementsCount,
            new IndirectComparator.AscendingIntComparator(fromNodes.buffer));
    //[[4 more lines in the extracted method]]
}
\end{lstlisting}
\end{subfigure}

\begin{subfigure}[t]{\linewidth}
\captionsetup{font=small}
\caption{\symboltpl\ From class com.carrotsearch.hppc.sorting.IndirectSort}
\label{fig:path-context-target}
\begin{lstlisting}[style=customjava, escapechar=!, linewidth=\linewidth]
public static int[] !\colorbox{green}{mergesort}!(int start, int length, IntBinaryOperator comparator) {
    final int[] src = createOrderArray(start, length);
    return mergesort(src, comparator);
}
\end{lstlisting}
\end{subfigure}
\label{fig:example-case-methods}
\end{figure}

\begin{figure}[t]
\captionsetup{font=small}
\caption{Extra \snippets extracted by \toolName as call path context. These details are required by the \testGenEngine to correctly invoke \symbole during the \harness generation process. Only the \snippets relevant to the example generated \harnesses are shown here. The full example, including the full prompt and all execution logs, is available on \href{https://github.com/sparkrew/fika/blob/main/Example.md}{GitHub}.}
\label{fig:example-case-extra}
\centering

\begin{subfigure}[t]{\linewidth}
\captionsetup{font=small}
\caption{Imports}
\begin{lstlisting}[style=customjava, escapechar=!, linewidth=\linewidth]
com.graphhopper.routing.ch.CHPreparationGraph.TurnCostFunction
\end{lstlisting}
\end{subfigure}

\begin{subfigure}[t]{\linewidth}
\captionsetup{font=small}
\caption{Factory methods / Constructors}
\begin{lstlisting}[style=customjava, escapechar=!, linewidth=\linewidth]
public static CHPreparationGraph edgeBased(int nodes, int edges, TurnCostFunction turnCostFunction) {
    return new CHPreparationGraph(nodes, edges, true, turnCostFunction);
}
\end{lstlisting}
\end{subfigure}

\begin{subfigure}[t]{\linewidth}
\captionsetup{font=small}
\caption{Setters}
\begin{lstlisting}[style=customjava, escapechar=!, linewidth=\linewidth]
public void addEdge(int from, int to, int edge, double weightFwd, double weightBwd) {
//[[11 more lines in the extracted call path context]]
}
\end{lstlisting}
\end{subfigure}

\end{figure}

\begin{figure}[]
\captionsetup{font=small}
\caption{An example generated \harness by \toolName that proves the dynamic reachability of a \dependency call site}. 
\label{fig:example-test}
\begin{lstlisting}[style=customjava, escapechar=!]
package com.graphhopper.routing.ch;
import com.graphhopper.routing.ch.CHPreparationGraph.TurnCostFunction;
import org.junit.jupiter.api.Test;

public class FikaTest {
    @Test
    public void testMergesort() {
        // Create an edge-based CHPreparationGraph to ensure origGraphBuilder is not null
        TurnCostFunction turnCostFunction = (in, via, out) -> 0;
        CHPreparationGraph graph = CHPreparationGraph.edgeBased(5, 10, turnCostFunction);
        // Add at least one edge to ensure the builder has data to sort
        graph.addEdge(0, 1, 0, 1.0, 1.0);
        // This call should traverse through origGraphBuilder.build() 
        // and eventually invoke IndirectSort.mergesort
        graph.!{\colorbox{yellow}{prepareForContraction();}}!
    }
}
\end{lstlisting}
\end{figure}

\subsection{Implementation}
We implement \toolName in Java as a proof of concept.
The implementation of all components in \toolName is publicly available at \href{https://github.com/sparkrew/fika}{https://github.com/sparkrew/fika}. In this section, we describe the design decisions behind the implementation of \toolName. 

\subsubsection{Static analysis}
We generate the call graph with \textit{SootUp} \unskip \unskip ~\cite{sootup2024}. Once the call graph is generated, we identify the \dependency call paths and the methods involved along these call paths. However, these method names are in bytecode, and when extracting context information for the path from the source code, we, need to map bytecode method names to source level method names. For this, we use \textit{Spoon}\unskip ~\cite{Pawlakr2016spoon} to query and extract the source code \href{https://github.com/sparkrew/fika/blob/main/api-finder/src/main/java/io/github/sparkrew/fika/api_finder/SourceCodeExtractor.java}{programmatically} by building the Abstract Syntax Tree (AST) of the project. 
%We use signature-based mapping criteria to accurately extract the corresponding sources:  if the method name contains ${<init>}$, we extract the corresponding constructor implementations from the source code; if the method name contains ${<clinit>}$, we map it to the static initializer blocks and and static field initializations of the class. If the method represents a regular method, we map it using its fully qualified signature. In all cases, we consider the method parameter types to correctly handle method overloading.  We also normalize names to handle encodings introduced by the compiler. Specifically we normalize the symbol \textit{\$} to \textit{.} to reduce false mismatches between bytecode naming and source naming. This is particularly useful when handling nested classes where bytecode and source may use different separators. If we cannot find a matching source method, we conservatively do not retrieve any context information for that method. 

\subsubsection{Dynamic analysis}
For dynamic analysis we determine call site coverage with the coverage reports generated by JaCoCo. To check whether the \dependency call sites of interest are covered by the executed tests, we parse the JaCoCo reports and verify whether the line at which the \dependency call site is invoked is marked as covered.

\subsubsection{LLM-based \harness generation}

% LLM
\toolName uses \model as the LLM for the \harness synthesis backend. We use \model as it is one of the strongest LLMs for code generation at the time of the implementation of \toolName. \toolName accesses the \model via its API \cite{deepseekai2025deepseekv32pushingfrontieropen}. 
The iterative \harness generation and validation workflow is orchestrated using LangChain\unskip ~\cite{langchain2023} and LangGraph \unskip ~\cite{langgraph2024}. 
As the unit testing frameworks for implementing and executing the generated \harnesses, we use \texttt{JUnit5} together with the \texttt{Surefire Maven Plugin}.

\section{Evaluation Methodology}
In this section, we present our dataset of open source Java projects, and then introduce the research questions that structure the empirical evaluation of \toolName.
The first two research questions correspond to the outputs of \toolName, and the third research question evaluates the LLM-based \harness generation pipeline. The final research question explores how \toolName improves state-of-the-art reachability analysis tools and helps developers prioritize and manage their dependencies effectively.

%%%%%%%%%%%%%%%%%%%%%%%

\subsection{Dataset}

\begin{table}[t]
\centering
\scriptsize
\captionsetup{font=small}
\caption{Projects used in our first part of evaluation along with their selected modules. We provide links to the GitHub repositories, commit SHAs, lines of code, the number of test cases, test coverage, the number of unique \symboltpl in each project, and the number of direct dependencies with compile or provided scope. The total number of direct and indirect dependencies, including both compile and provided scopes, is given in parentheses.}
\label{tab:project-commits}
\setlength{\tabcolsep}{4pt}
\begin{tabular}{|l|l|r|r|r|r|r|r|r|}
\hline
% ---- Headers ----
Project &
Module &
\makecell[r]{Commit \\ SHA} &
LoCs &
Tests &
Cov. &
\makecell[r]{\textsc{\symboltpl}} &
\makecell[r]{Dir. deps \\ (All deps)} \\
\hline

\rowcolor{rowGray}
\texttt{\projFlink} & 
\texttt{\nbModuleFlink} & 
\nbCommitFlink & 
\nbLocFlink & 
\nbTestsFlink & 
\nbCoverageFlink & 
\nbUniqueTPLMethods{109} &
\nbDepsFlink      \\

\texttt{\projGraphhopper} & 
\texttt{\nbModuleGraphhopper} & 
\nbCommitGraphhopper & 
\nbLocGraphhopper & 
\nbTestsGraphhopper& 
\nbCoverageGraphhopper & 
\nbUniqueTPLMethods{276} & 
\nbDepsGraphhopper \\

\rowcolor{rowGray}
\texttt{\projJooby} & 
\texttt{\nbModuleJooby} & 
\nbCommitJooby & 
\nbLocJooby & 
\nbTestsJooby & 
\nbCoverageJooby & 
\nbUniqueTPLMethods{33} & 
\nbDepsJooby     \\

\texttt{\projMybatis} & 
\texttt{\nbModuleMyBatis} & 
\nbCommitMyBatis & 
\nbLocMyBatis & 
\nbTestsMyBatis & 
\nbCoverageMyBatis & 
\nbUniqueTPLMethods{35} & 
\nbDepsMyBatis    \\

\rowcolor{rowGray}
\texttt{\projPdfbox} & 
\texttt{\nbModulePdfBox} & 
\nbCommitPdfBox & 
\nbLocPdfBox & 
\nbTestsPdfBox & 
\nbCoveragePdfBox & 
\nbUniqueTPLMethods{49} & 
\nbDepsPdfBox     \\

\texttt{\projTablesaw} & 
\texttt{\nbModuleTablesaw} & 
\nbCommitTablesaw & 
\nbLocTablesaw & 
\nbTestsTablesaw & 
\nbCoverageTablesaw & 
\nbUniqueTPLMethods{26} & 
\nbDepsTablesaw   \\

\rowcolor{rowGray}
\texttt{\projTika} & 
\texttt{\nbModuleTika} & 
\nbCommitTika & 
\nbLocTika & 
\nbTestsTika & 
\nbCoverageTika & 
\nbUniqueTPLMethods{23} & 
\nbDepsTika       \\

\texttt{\projPoiTl} & 
\texttt{\nbModulePoiTl} & 
\nbCommitPoiTl & 
\nbLocPoiTl & 
\nbTestsPoiTl & 
\nbCoveragePoiTl & 
\nbUniqueTPLMethods{812} & 
\nbDepsPoiTl      \\

\hline
\end{tabular}
\end{table}

We analyze a set of public GitHub projects using \toolName and evaluate its ability to find evidence for the \executability of \dependency call sites.
We start from the set of 30 open-source Java Maven projects collected by Soto-Valero \etal \cite{soto2023specialize}. \modified{This dataset was assembled from two sources: popular Maven projects with passing builds from an existing benchmark \cite{durieux2021duets}, and top-starred GitHub Java repositories. The projects were filtered to require a successful Maven build, at least one compile-scoped \dependency, and an executable test suite. We chose this dataset because its selection criteria match the requirements of our study.}
From these projects, we remove two projects whose current versions have migrated from Maven to Gradle.
Then, we select the projects that use JUnit5 and filter out five projects that depend on TestNG or JUnit4. We also discard nine projects that depend on Java versions lower than 17. This leaves 14 projects. Then, we compile and execute their tests. If the execution is unsuccessful, we check out to the most recent tag of the project as tagged commits indicate a more stable update. If that version also fails to build and run the tests, we discard that project. We discard a total of five projects due to such build failures.
We discard one additional project, \texttt{checkstyle}, because it is a meta-level static analysis tool. 
% It includes source files generated during the Maven \textit{generate-sources} phase, which makes it out of scope for \toolName, as \toolName currently does not consider source code under generated sources.
Finally, we select \nbprojects projects that use JUnit5 and successfully build with Java 17 or higher at the time of experimenting. 

Statistics about the selected \nbprojects projects to answer the first three research questions are given in \autoref{tab:project-commits}.
The second column shows the module we analyze when the project is a multi module project. We select the same modules as Soto Valero \etal \cite{soto2023specialize}.
The third column provides the commit hash and a link to the analyzed version of each project.
The next column reports the number of lines of Java code (LoC) according to the Unix \texttt{cloc} command, which is an indicator of the size of the projects. The following two columns give the number of test cases executed by the Maven surefire plugin, and the test coverage reported by the JaCoCo plugin. 
\modified{The last two columns provide the number of unique \dependency methods (\symboltpl), and the number of direct compile and provided scoped \dependencies as resolved in the Maven dependency tree. 
We do not consider sub-modules from the same project as \dependencies.
The sum of both direct and indirect \dependencies is shown in brackets. 
The selected projects vary in size, number of tests, test coverage, and the number of \dependencies as detailed in the table.}
% The maximum number of direct \dependencies (10) is observed in \texttt{graphhopper-core}. In terms of indirect \dependencies, some projects with a lower number of direct \dependencies are resolved with a higher number of indirect \dependencies. For example, \texttt{poi-tl} has a total of 36 \dependencies, although only five of them are direct.

We use these 8 modules to answer the first three research questions. For the final research question, we use all the modules from the 8 projects which we explain in detail in the next section.

\subsection{Protocol}
\label{sec:protocol}

We answer the following four research questions in order to evaluate \toolName.

\newcommand\rqstatic{To what extent can \toolName successfully extract static call paths to reach \dependency call sites? \xspace}

\newcommand\rqtests{To what extent do developer-written tests provide dynamic guarantees of \dependency executability? \xspace}

\newcommand\rqfika{To what extent can \toolName improve dynamic guarantees of \dependency executability? \xspace}

\newcommand\rqllms{How do \toolName's design decisions impact the effectiveness of the generated \harnesses? \xspace}

\newcommand\rqsemgrep{How can \toolName's executability analysis improve state-of-the-art \dependency reachability analysis tools? \xspace}

% \begin{enumerate}[label=RQ\arabic*:, ref=RQ\arabic*]

%     \item \label{rq:static} [static] \textbf{\rqstatic}
%     \item \label{rq:tests} [dynamic] \textbf{\rqtests}
%     \item \label{rq:fika} [dynamic with generated tests] \textbf{\rqfika}
%     \item \label{rq:prompts} [LLM-based test generation] \textbf{\rqllms}
% \end{enumerate}

\textbf{\textbf{RQ1: }\textbf{\rqtests}}
In RQ1, our goal is to look at how many \dependency call sites within a project are verified as \executable after running the developer-written test suites. 

To answer this question, we first run \toolName to generate the call graph and identify all \dependency call sites that are statically reachable from a \symbole. 
Then, using \toolName, we check how many of the identified call sites are covered by the current project tests. 
To allow the collection of test coverage, we add JaCoCo maven plugin, if it is not already present in the projects. 

We report the number of \dependency call sites that are effectively triggered, as evidenced by the existing project test suites. 
Doing this, we also establish a baseline that \textsc{fika}’s \testGenEngine can improve upon.

\textbf{\textbf{RQ2: }\textbf{\rqfika}}

With this RQ, we evaluate the ability of \toolName to perform a targeted \harness generation to dynamically reach \dependency call sites.
Here, using \toolName, we attempt to provide evidence for the \executability of \dependency call sites that are not invoked by any test in the project’s test suite. This demonstrates the added value of \toolName.

At this stage, \toolName has a list of call paths to reach \dependency call sites from public project methods.
We filter out the call paths  to call sites that are already covered by the existing test suite.
We sort the remaining paths according to their length. The shorter paths are selected first for \harness generation, as path length is an indication of the potential difficulty of generating a \harness.
Then, we generate \harnesses with \toolName, for one path at a time, following the sorted order.
Recall that each call path has a unique combination of \symbole, \symboldp, and \symboltpl.
When a successful \harness is generated for a call site (a combination of \symboldp, and \symboltpl), if the same call site appears again with a different \symbole, \toolName proceeds to the next candidate, as the goal is to collect evidence for the \executability of that \dependency call site, not to have multiple \harnesses.

After this process, we calculate the number of successful \harnesses generated by \toolName and report the improved evidence of \executability for \dependency call sites. A key indicator here is the number of additional \dependency call sites for which we have evidence of \executability, compared to the original test suites.
% For the call sites for which a test is generated, we can confirm their reachability with certainty. However, the call sites for which \toolName fails to generate a test remain unconfirmed in terms of reachability.

\textbf{\textbf{RQ3: }\textbf{\rqllms}}

\modified{In RQ3, we conduct an incremental ablation study to analyze four aspects of the proposed approach: (i) the impact of providing path-related data retrieved from static analysis, (ii) the impact of adding entry-point-related context alongside the path information, (iii) the impact of selecting the shortest path to reach a \dependency method, and (iv) the impact of the feedback mechanism.}  
%We denote the two baseline configurations as \textbf{BL1} and \textbf{BL2} (BL = baseline), and refer to the complete approach as \textbf{FIKA}.

In the first configuration (BL1), we create a baseline prompt that provides only the project method containing the target call site (\symboldp) to the LLM, without including the path to a public entry method. This setting allows us to evaluate whether the static analysis step, which identifies paths to reach \dependency call sites, actually improves the \harness generation process.

In the second configuration (BL2), the prompt includes both the path and the implementations of all methods along that path. However, we do not provide additional entry point related context, such as constructors, factory methods, or field declarations of the class to which the entry point belongs. This configuration allows us to evaluate whether providing this additional contextual information improves the LLM’s ability to generate successful \harnesses.

\modified{In the third ablation (BL3), instead of choosing the shortest path to reach a \dependency call site, we extract a random path from a random entry point. We extract the same information as in the original approach, including the context related to the entry point, and prompt the LLM. With this configuration, our goal is to analyze how the path selection algorithm of \toolName impacts \harness generation success.}

In the fourth configuration, which corresponds to the complete approach, we evaluate the feedback mechanism and verify whether the errors or constraint violations from previous attempts can guide the LLM toward successful \harness generation.

\textbf{\textbf{RQ4: }\textbf{\rqsemgrep}}

% \modified{
To evaluate \toolName against state of the art approaches for reachability analysis, we compare it with one static reachability analysis tool and one dynamic analysis tool. Specifically, we compare the static analysis step of \toolName with a pattern based reachability analyzer, and the dynamic analysis step with a coverage guided directed fuzzer.

We perform the comparison on the real-world problem of vulnerability reachability. Reachability analysis is commonly used to reduce false positive vulnerabilities by determining whether vulnerable functions in dependencies can actually be reached.
To identify vulnerable methods within \dependencies, we use \semgrep \footnote{https://semgrep.dev}, an open source, state of the art static application security testing tool. One of the main features of \semgrep is to detect security issues introduced by \dependencies. 

\modified{As case studies for the vulnerability analysis, we select the eight projects listed in \autoref{tab:project-commits} and consider all of their Maven modules. This broad scope also allows us to validate \toolName on modules beyond the initial dataset.} We first run \semgrep on all modules at the commits listed in \autoref{tab:project-commits}. Under these commits, \semgrep reports vulnerabilities only in \texttt{poi-tl}. Therefore, for the remaining projects, we go further back in the history and check out the oldest commit that can be successfully compiled and executed with Java 17+ and JUnit 5. We choose the oldest executable commit because it increases the likelihood of finding outdated and potentially vulnerable \dependencies. In this way, we identify 12 additional modules with vulnerabilities, resulting in 13 modules in total.

We use the vulnerability information reported by \semgrep as the starting point for the reachability analysis and then analyze whether the identified vulnerable \dependency methods can actually be reached.
We divide the vulnerability reachability analysis into two parts: a) comparison of the static analysis of \toolName with pattern based reachability analysis, and b) comparison of the dynamic analysis of \toolName with a directed fuzzer.

\paragraph*{RQ4.a: Comparison of \toolName's static analysis with pattern based reachability analysis}

As the baseline for static reachability analysis, we use the pattern based reachability analysis provided by \semgrep. \semgrep has also been used as a baseline in previous studies of vulnerability analysis \cite{liu2023qatools, li2025mcpsemgrep, Guo2024OutsideTCA, grgic2026propagation}. %https://asta.allen.ai/chat/80f512d5-2d96-42ac-9425-21f3f3f598a8 papers that use semgrep as part of their experiments

\semgrep's pattern based reachability analysis classifies a Common Vulnerabilities and Exposures (CVE) in a \dependency as either \textit{reachable}, \textit{undetermined}, or \textit{unreachable}. This analysis is based on a set of static rules. A reachability rule can be manually defined for a CVE and consists of patterns that identify the use of vulnerable methods originating from \dependencies. The CVE is marked as \textit{reachable} if the pattern is matched and \textit{unreachable} otherwise.

Some rules do not define a pattern and instead only look for the presence of the \dependency in the dependency tree. In these cases, \semgrep marks the CVE as either \textit{reachable} or \textit{undetermined}. The exact semantics of these classifications are not publicly documented.

We run the static analysis step of \toolName on the 13 module versions with at least one vulnerability according to \semgrep. For cases where \semgrep defines a pattern, we use the same pattern to identify the corresponding vulnerable \symboltpl and analyze its static reachability with \toolName. For cases where \semgrep does not define a pattern, we check the static reachability of any method originating from the vulnerable \dependency. We record whether each vulnerable \symboltpl reported by \semgrep as \textit{reachable} or \textit{undetermined} is present in the code base and whether \toolName's static analysis identifies a path to reach it.

\modified{
\paragraph*{RQ4.b: Comparison of \toolName's dynamic analysis with fuzzing}

As the baseline for dynamic reachability (\executability) analysis, we use \fuzzer \cite{dechand2026jazzer}, an open source coverage guided directed fuzzer that supports Java. \fuzzer allows us to specify methods to instrument and target for coverage. We therefore configure \fuzzer to maximize coverage toward the target call sites and their entry points identified by the static analysis step of \toolName.
The goal is to provide \fuzzer and the \testGenEngine of \toolName with the same target information before performing dynamic analysis. 
% Both approaches therefore attempt to reach the same vulnerable methods identified by the static analysis. 

For each target  call site, we first run \toolName's \harness generation and record the time required to generate a successful \harness. We then give \fuzzer the same amount of time to reach the same target. This gives both approaches the same time budget for \executability analysis.

We compare the successful \harnesses generated by \toolName with the inputs found by \fuzzer that reach the target vulnerable methods. For \toolName, a successful result is a \harness that can be executed and reaches the target vulnerable method. For \fuzzer, a successful result is an input that reaches the corresponding target during fuzzing. 
}
%%%%%%%%%%%%%%%%%%%%%%%%%%%%%%%%%%%%%%%%%%%%%%%%
%%%%%%%%%%%%%%%%%%%%%%%%%%%%%%%%%%%%%%%%%%%%%%%%
%%%%%%%%%%%%%%%%%%%%%%%%%%%%%%%%%%%%%%%%%%%%%%%%
%%%%%%%%%%%%%%%%%%%%%%%%%%%%%%%%%%%%%%%%%%%%%%%%
\section{Experimental results}
In this section, we present the results from the evaluation of \toolName with the selected eight projects and using the protocol listed in \autoref{sec:protocol}.
A summary of the answers to our first three research questions is presented in \autoref{tab:rq123}, and the answer to the final research question is presented in \autoref{tab:rq5} and \autoref{tab:rq5_42}.

\begin{table*}
\scriptsize
\centering
\captionsetup{font=small}
\caption{Results from the evaluation of \toolName on the first three research questions.
\textsc{RQ1} relates to the execution of existing tests. \textsc{TPL call sites} gives the number of call sites. \textsc{TPLs executed by the tests} presents the total number of call sites, that have the guarantees of \executability through developer-written tests.
\textsc{RQ2} reflects the ability of \toolName to improve the \executability guarantees. \textsc{TPLs executed by fika-generated scenarios} presents the call sites covered by new \harnesses as a fraction of those for which generation is attempted. 
The next two columns, \textsc{Total TPLs with reachability guarantees} and \textsc{Additional guarantees by \toolName}, present the total number of call sites that have guarantees of \executability and the improvement achieved by \toolName.
\textsc{RQ3} presents a comparison of \toolName’s design decisions, where each column shows the successful cases generated with the baselines (BL1: no path details, BL2: path details without entry-point context, BL3: random path selection) and across iterations (I1–I5).}
\label{tab:rq123}
\setlength\tabcolsep{5.4pt}
\def\arraystretch{1.2}
\begin{tabular}{|l||r|>{\raggedleft\arraybackslash}p{7.1em}||>{\raggedleft\arraybackslash}p{7.9em}|>{\raggedleft\arraybackslash}p{7.1em}|>{\raggedleft\arraybackslash}p{5.9em}||r|r|r|r|r|r|r|r|}
\hline

\multirow{2}{*}{\textsc{Project}} &
\multicolumn{2}{c||}{\cellcolor[rgb]{0.988,0.996,0.922}\textsc{\textbf{RQ1}}} &
\multicolumn{3}{c||}{\cellcolor[rgb]{0.886,0.992,0.988}\textsc{\textbf{RQ2}}} &
\multicolumn{8}{c|}{\cellcolor[rgb]{0.95,0.95,1}\textsc{\textbf{RQ3}}} \\

\hhline{|~||--||---||--------|}

&
\cellcolor[rgb]{0.988,0.996,0.922}\textsc{\makecell[r]{TPL\\call \\ sites}} &
\cellcolor[rgb]{0.988,0.996,0.922}\textsc{\makecell[r]{TPLs \\ executed \\ by the tests}} &
\cellcolor[rgb]{0.886,0.992,0.988}\textsc{\makecell[r]{TPLs \\ executed by \\ fika-generated \\ scenarios}} &
\cellcolor[rgb]{0.886,0.992,0.988}\textsc{\makecell[r]{Total \\ TPLs with \\reachability \\ guarantees}} &
\cellcolor[rgb]{0.886,0.992,0.988}\textsc{\makecell[r]{Additional \\ guarantees \\ by Fika}} &
\cellcolor[rgb]{0.95,0.95,1}\textsc{BL1} &
\cellcolor[rgb]{0.95,0.95,1}\textsc{BL2} &
\cellcolor[rgb]{0.95,0.95,1}\textsc{BL3} &
\cellcolor[rgb]{0.95,0.95,1}\textsc{I1} &
\cellcolor[rgb]{0.95,0.95,1}\textsc{I2} &
\cellcolor[rgb]{0.95,0.95,1}\textsc{I3} &
\cellcolor[rgb]{0.95,0.95,1}\textsc{I4} &
\cellcolor[rgb]{0.95,0.95,1}\textsc{I5} \\

\hline
%%%%%%%%%%%%%%%%%%%%%%%%%%%%%%%%%%%%%%%%%%%%%%%%%%%%
\rowcolor{rowGray}
\texttt{\projFlink} &
\nbReachableCallSites{143} &
\nbAllCoveredByExistingTests{46/143 (32\%)} &
\nbCoveredByNewTests{47/97 (48\%)} &
\nbCoveredByTotal{93/143 (65\%)} &
\nbIncrease{33\%} &
\nbWeakest{12} &
\nbLessWeak{28} &
\nbRandom{15} &
\nbIun{28} &
\nbIdeux{39} &
\nbItrois{43} &
\nbIquatre{44} &
\nbIcinq{47} \\

%%%%%%%%%%%%%%%%%%%%%%%%%%%%%%%%%%%%%%%%%%%%%%%%%%%%
\texttt{\projGraphhopper} &
\nbReachableCallSites{668} &
\nbAllCoveredByExistingTests{469/668 (70\%)} &
\nbCoveredByNewTests{77/199 (39\%)} &
\nbCoveredByTotal{546/668 (82\%)} &
\nbIncrease{12\%} &
\nbWeakest{12} &
\nbLessWeak{41} &
\nbRandom{47} &
\nbIun{53} &
\nbIdeux{64} &
\nbItrois{70} &
\nbIquatre{77} &
\nbIcinq{77} \\

%%%%%%%%%%%%%%%%%%%%%%%%%%%%%%%%%%%%%%%%%%%%%%%%%%%%
\rowcolor{rowGray}
\texttt{\projJooby} &
\nbReachableCallSites{52} &
\nbAllCoveredByExistingTests{22/52 (42\%)} &
\nbCoveredByNewTests{28/30 (93\%)} &
\nbCoveredByTotal{50/52 (96\%)} &
\nbIncrease{54\%} &
\nbWeakest{11} &
\nbLessWeak{13} &
\nbRandom{18} &
\nbIun{19} &
\nbIdeux{26} &
\nbItrois{27} &
\nbIquatre{28} &
\nbIcinq{28} \\

%%%%%%%%%%%%%%%%%%%%%%%%%%%%%%%%%%%%%%%%%%%%%%%%%%%%
\texttt{\projMybatis} &
\nbReachableCallSites{49} &
\nbAllCoveredByExistingTests{19/49 (39\%)} &
\nbCoveredByNewTests{23/30 (77\%)} &
\nbCoveredByTotal{42/49 (86\%)} &
\nbIncrease{47\%} &
\nbWeakest{7} &
\nbLessWeak{11} &
\nbRandom{9} &
\nbIun{23} &
\nbIdeux{23} &
\nbItrois{23} &
\nbIquatre{23} &
\nbIcinq{23} \\

%%%%%%%%%%%%%%%%%%%%%%%%%%%%%%%%%%%%%%%%%%%%%%%%%%%%
\rowcolor{rowGray}
\texttt{\projPdfbox} &
\nbReachableCallSites{591} &
\nbAllCoveredByExistingTests{108/591 (18\%)} &
\nbCoveredByNewTests{223/483 (46\%)} &
\nbCoveredByTotal{331/591 (56\%)} &
\nbIncrease{38\%} &
\nbWeakest{123} &
\nbLessWeak{146} &
\nbRandom{73} &
\nbIun{153} &
\nbIdeux{188} &
\nbItrois{199} &
\nbIquatre{212} &
\nbIcinq{223} \\

%%%%%%%%%%%%%%%%%%%%%%%%%%%%%%%%%%%%%%%%%%%%%%%%%%%%
\texttt{\projTablesaw} &
\nbReachableCallSites{50} &
\nbAllCoveredByExistingTests{35/50 (70\%)} &
\nbCoveredByNewTests{7/15 (47\%)} &
\nbCoveredByTotal{42/50 (84\%)} &
\nbIncrease{14\%} &
\nbWeakest{2} &
\nbLessWeak{3} &
\nbRandom{4} &
\nbIun{5} &
\nbIdeux{5} &
\nbItrois{5} &
\nbIquatre{6} &
\nbIcinq{7} \\

%%%%%%%%%%%%%%%%%%%%%%%%%%%%%%%%%%%%%%%%%%%%%%%%%%%%
\rowcolor{rowGray}
\texttt{\projTika} &
\nbReachableCallSites{30} &
\nbAllCoveredByExistingTests{8/30 (27\%)} &
\nbCoveredByNewTests{15/22 (68\%)} &
\nbCoveredByTotal{23/30 (77\%)} &
\nbIncrease{50\%} &
\nbWeakest{2} &
\nbLessWeak{8} &
\nbRandom{9} &
\nbIun{13} &
\nbIdeux{14} &
\nbItrois{15} &
\nbIquatre{15} &
\nbIcinq{15} \\

%%%%%%%%%%%%%%%%%%%%%%%%%%%%%%%%%%%%%%%%%%%%%%%%%%%%
\texttt{\projPoiTl} &
\nbReachableCallSites{1636} &
\nbAllCoveredByExistingTests{1047/1636 (64\%)} &
\nbCoveredByNewTests{189/589 (32\%)} &
\nbCoveredByTotal{1236/1636 (76\%)} &
\nbIncrease{12\%} &
\nbWeakest{52} &
\nbLessWeak{137} &
\nbRandom{62} &
\nbIun{90} &
\nbIdeux{144} &
\nbItrois{159} &
\nbIquatre{178} &
\nbIcinq{189} \\

\hline
Total &
\nbReachableCallSites{3219} &
\nbAllCoveredByExistingTests{1754/3219 (54\%)} &
\nbCoveredByNewTests{609/1465 (42\%)} &
\nbCoveredByTotal{2363/3219 (73\%)} &
\nbIncrease{19\%} &
\nbWeakest{221} &
\nbLessWeak{387} &
\nbRandom{237} &
\nbIun{384} &
\nbIdeux{503} &
\nbItrois{541} &
\nbIquatre{583} &
\nbIcinq{609} \\
%%%%%%%%%%%%%%%%%%%%%%%%%%%%%%%%%%%%%%

\hline
\end{tabular}
\end{table*}

\subsection*{\textbf{RQ1: \rqtests}}

With this research question, we analyze how many \dependency call sites are proven to be \executable based on the execution of existing developer-written project test suites. The results are presented in the RQ1 column of \autoref{tab:rq123}. 

The second column, \textit{TPL call sites}, shows the number of call sites that are reachable from public methods of the project through static analysis. The third column reports the call sites that are shown to be \executable through developer-written tests.

As shown in \autoref{tab:project-commits}, almost all projects in our dataset have more than 100 developer-written test cases.
These test suites cover some, but not all, \dependency call sites.
For example, in \texttt{graphhopper}, 276 unique dependency methods are called within the project code. These 276 methods appear across 668 call sites, since a single \symboltpl may be invoked by multiple \symboldp. Out of the 668 call sites, 469 (70\%) are covered by the existing tests.
The least covered project is \texttt{pdfbox}, where only 108 out of 591 (18\%) call sites are covered, even though the overall test coverage of the project is 60\%. This is because in \texttt{pdfbox}, the most frequently used \dependency calls are related to logging from the \path{org.apache.\allowbreak commons.logging} library and these logging call sites are often executed in situations that normal execution does not reach, such as edge cases where errors are thrown or when debugging is needed. 

Developer-written tests generally cover \dependency call sites under three scenarios. First, developers recognize that a \dependency method’s behavior directly impacts the client project and therefore write test cases that explicitly test its behavior. For example, in the project \texttt{Tika}, the class \texttt{TikaInputStream} extends \texttt{TaggedInputStream} from the \dependency\xspace \texttt{commons-io\allowbreak:commons-io}. Within this class, the method \texttt{mark} overrides the corresponding method in the superclass and also explicitly invokes it. Since the behavior of \texttt{mark} is an essential part of the overall program behavior, the developers of \texttt{Tika} have added nine tests in the test class \texttt{TikaInputStreamTest} that invoke \texttt{mark} and include assertions to validate its behavior. 

Second, developers do not target a specific \dependency call site but instead test its direct caller, and as a result, the \dependency call site is also invoked. In this case, no assertion directly targets the \symboltpl, but the test may still depend on its behavior. For example, in \texttt{Tika}, the \symboldp\xspace \texttt{seekTo} method calls the \symboltpl\xspace \texttt{skipFully} from the \dependency\xspace \texttt{commons-io\allowbreak:commons-io}. Three developer-written test cases directly invoke and validate the behavior of \texttt{seekTo}, and as a result, \texttt{skipFully} is also invoked, and is indirectly tested.

Third, developers write tests for a method that  triggers the \dependency call site without intending to test the behavior of the \symboltpl or its direct caller. These tests still expose \dependency interactions with the project, although they do not explicitly evaluate their correctness.
For example, the public method \texttt{extractRootElement} (\symboldp) in the class \texttt{XmlRootExtractor} in \texttt{Tika} invokes the constructor \texttt{UnsynchronizedByteArrayInputStream} (\symboltpl) from the \dependency\xspace \texttt{commons-io\allowbreak:commons-io}. There is no dedicated test class for \texttt{XmlRootExtractor}, and no test explicitly calls the \symboldp. However, the method \texttt{getMimeType} in the class \texttt{MimeTypes} invokes the \symboldp and is covered by two developer-written tests. 
% \texttt{testMSAccessByName} and \texttt{testZipXFiles}. 
Thus, in this case, the \symboltpl also gets covered, even though it is not explicitly intended by the developers.

Overall, good developer-written test suites in Java projects naturally trigger \dependency call sites. Yet, only \testcov of call sites across these well-tested projects are covered by the existing tests, highlighting the need for additional \harnesses to confirm their executability, as well as the need for \toolName. 

\begin{tcolorbox}[boxrule=1pt,arc=.3em, left=2pt, right=2pt]
  \textbf{Answer to RQ1}: Developer-written test suites confirm the executability of 18\% and up to 70\% of \dependency call sites, depending on the project. On the one hand, this supports our idea of getting  guarantees of \executability through test execution. On the other hand, it highlights the need for an automated tool like \toolName to strengthen the level of such guarantees.
\end{tcolorbox}

\subsection*{\textbf{RQ2: \rqfika}}

With RQ2, we evaluate \toolName's ability to improve the executability guarantees of \dependencies by generating new \harnesses. The \harness generation focuses on the call sites which are not covered by the existing tests, hereby providing value. We present the results in the column \textsc{RQ2} of \autoref{tab:rq123}.

\begin{figure}[]
\captionsetup{font=small}
\caption{An example generated \harness by \toolName that goes through a call chain of length 8 to reach a call site with \symboltpl, \texttt{Log.debug()} that is not covered by developer-written tests. The comments within [[]] are not part of the original \harness. They are included only to keep the example succinct.} 
\label{fig:example-test-rq3}
\begin{lstlisting}[style=customjava, escapechar=!]
package org.apache.pdfbox.pdfparser;
// [[5 imports in the actual reachability scenario]]

public class LogdebugFikaTest {
    @Test
    public void testParseToLogDebug() throws IOException {
        // [[3 comment lines]]
        String fdfContent = "%FDF-1.2\n" +
                // [[9 more lines]]
                "%%EOF";
        byte[] fdfBytes = fdfContent.getBytes();
        RandomAccessRead source = new RandomAccessReadBuffer(new ByteArrayInputStream(fdfBytes));
        FDFParser parser = new FDFParser(source);
        try {
            parser.parse();
        } catch (IOException e) {
            // Expected - the file is not a valid FDF, but we only care about
            // executing the call chain up to Log.debug()
}}}
\end{lstlisting}
\end{figure}

As shown in the column \textsc{TPLs executed by fika-generated scenarios}, \toolName generates successful new \harnesses for all projects, covering new call sites that are not exercised by the existing test suites.
For \texttt{pdfbox}, \toolName produces 223 successful \harnesses, covering 46\% of the attempted targets, which is more than twice as much as what the original test suite covers. They reach the error-related call sites involving \dependency methods such as \path{Log.error} and \path{Log.warn}, which are typically not triggered by developer-written tests, as discussed above.
For example, the \harness presented in \autoref{fig:example-test-rq3}, starts from the entry point \texttt{FDFParser\allowbreak.parse}, goes through 8 more hops to reach the \dependency call site with the \symboltpl\xspace, \texttt{org.apache.\allowbreak commons.logging.\allowbreak Log.debug}. 

\begin{table}
\scriptsize
\centering
\captionsetup{font=small}
\caption{Distribution of path lengths for generated \harnesses}
\label{tab:rq3paths}
\setlength\tabcolsep{15pt}

\begin{tabular}{|l||r|r|r|r|}

\hline

\multirow{2}{*}{\textsc{Project}} &
\multicolumn{4}{c|}{\textsc{Path length}} \\

\hhline{|~||----|}

&

\textsc{\makecell[r]{1}} &
\textsc{\makecell[r]{2}} &
\textsc{\makecell[r]{3}} &
\textsc{\makecell[r]{4+}} \\

\hline

\rowcolor{rowGray}
\texttt{\projFlink} &
\nbIdeux{27} &
\nbItrois{15} &
\nbIquatre{5} &
\nbIcinq{0} \\

\texttt{\projGraphhopper} &
\nbIdeux{62} &
\nbItrois{10} &
\nbIquatre{4} &
\nbIcinq{1} \\

\rowcolor{rowGray}
\texttt{\projJooby} &
\nbIdeux{26} &
\nbItrois{0} &
\nbIquatre{2} &
\nbIcinq{0} \\

\texttt{\projMybatis} &
\nbIdeux{21} &
\nbItrois{2} &
\nbIquatre{0} &
\nbIcinq{0} \\

\rowcolor{rowGray}
\texttt{\projPdfbox} &
\nbIdeux{73} &
\nbItrois{123} &
\nbIquatre{11} &
\nbIcinq{16} \\

\texttt{\projTablesaw} &
\nbIdeux{4} &
\nbItrois{3} &
\nbIquatre{0} &
\nbIcinq{0} \\

\rowcolor{rowGray}
\texttt{\projTika} &
\nbIdeux{14} &
\nbItrois{1} &
\nbIquatre{0} &
\nbIcinq{0} \\

\texttt{\projPoiTl} &
\nbIdeux{169} &
\nbItrois{18} &
\nbIquatre{1} &
\nbIcinq{1} \\

\hline

Total &
396 &
172 &
23 &
18 \\

\hline
\end{tabular}
\end{table}

% jooby
For \texttt{jooby}, \toolName provides \executability guarantees for 28 out of the 30 previously non-confirmed call sites, which is almost perfect. 
The remaining two call sites have non-trivial preconditions that are hard to satisfy, and \toolName fails to generate a \harness that meets these conditions. For example, the public method \href{https://github.com/jooby-project/jooby/blob/d2272e7bba0e9c6c6f97827eccd2550c56d3a1cf/jooby/src/main/java/io/jooby/Jooby.java#L664C3-L672C4}{\texttt{Jooby.getTmpdir}} may reach the \symboltpl\xspace \path{com.typesafe.\allowbreak config.Config.\allowbreak getString} only under the condition that (i) an internal cache is \texttt{null} and (ii) the \texttt{Environment} variable is correctly initialized, so that \path{getEnvironment().\allowbreak getConfig().\allowbreak getString} can execute. 
The current \testGenEngine does not manage to construct that state.

It is important that the generated \harnesses can trigger paths longer than one, as some call sites can only be reached through multiple hops.
In \autoref{tab:rq3paths}, we present the path lengths of the \harnesses generated by \toolName. The results show that \toolName can successfully generate \harnesses for call sites that are deep within the project code. In larger modules such as \texttt{graphhopper}, \texttt{pdfbox}, and \texttt{poi-tl}, 18 paths extend beyond four hops.
We also observe that the majority of paths have lengths of 1 or 2 (568 out of 609). This is because \toolName prioritizes shorter paths when they exist.%, as they are potentially easier to reach. 

\toolName significantly improves \dependency executability guarantees by generating new \harnesses. 
The additional evidence for \executability provided by \toolName, shown in the column \textsc{Additional guarantees by \toolName} of \autoref{tab:rq123}, exceeds 12\% for every project.
As a result, the total executability (\textsc{Total Guarantees}) exceeds 50\% for all projects after applying \toolName and averages 73\%.

\begin{tcolorbox}[boxrule=1pt,arc=.3em, left=2pt, right=2pt]
  \textbf{Answer to RQ2}: \toolName successfully generates new \harnesses that provide \executability guarantees for many previously non-covered \dependency call sites across all projects. It is capable of reaching call sites that, in some cases, require traversing paths of more than four method calls. The improvement achieved by \toolName ranges from a 12\% increase in \texttt{graphhopper} to 54\% in \texttt{jooby}.
\end{tcolorbox}

\subsection*{\textbf{RQ3: \rqllms}}

This research question evaluates how the prompt design in \toolName influences the results.
\modified{In \autoref{tab:rq123} (RQ3), we compare three baselines: \textsc{BL1} uses a prompt without path details, \textsc{BL2} includes path details but omits entry-point-related context, and \textsc{BL3} selects a random path from a random entry point instead of selecting the shortest path to reach the \dependency call site.} 
We also show the results of the full configuration across feedback iterations (\textsc{I1}–\textsc{I5}).

% the cumulative number of successful test generations at each iteration.

For all projects, the configuration BL1 shows the lowest performance. Without information about the path, the LLM has to generate a \harness for the \symboldp without knowing whether it is directly invocable. Consequently, many generated \harnesses, including those that execute successfully, rely on reflection to access private fields and methods, which makes the generated \harnesses very brittle.
Adding path information in BL2 consistently improves performance across all projects. For example, the number of successful \harnesses increases from 12 to 41 for \texttt{\projGraphhopper}, and from 52 to 137 for \texttt{\projPoiTl}. This demonstrates that \toolName’s static analysis phase improves the quality of the generated \harnesses by constraining the LLM to follow a feasible invocation path rather than guessing access patterns. 

Compared to the full prompt with all contextual information, BL2 achieves lower performance for most projects. The only exception is \texttt{poi-tl}, where BL2 slightly exceeds the first iteration of \toolName (137 vs. 90). This suggests that the LLM can succeed when provided with limited context, whereas in most projects the additional entry-point-related context in the full prompt provides an immediate benefit.

\modified{BL3 further evaluates the impact of the path selection strategy.  Compared to the shortest path selection strategy, BL3 performs worse for all projects. For example, the number of successful \harnesses increases from 47 to 53 for \texttt{\projGraphhopper}, and from 73 to 153 for \texttt{\projPdfbox} between BL3 and I1. Selecting the shortest path does increase the chances of generating more successful \harnesses.}

The second part of the RQ3 section of \autoref{tab:rq123} shows that the feedback loop of \testGenEngine is effective. For most projects, the number of successful \harness generation attempts increases across iterations (e.g.  \texttt{pdfbox}: 153$\rightarrow$223; \texttt{poi-tl}: 90$\rightarrow$189), indicating that iterative refinement helps the model recover from constraint violations, compilation errors, and execution errors. However, \texttt{mybatis-3} shows no improvement after the first iteration (23 in \textsc{I1} and 23 in \textsc{I5}).
When investigating \texttt{mybatis-3}, we find that the failed cases primarily involve call paths with methods that require complex objects as parameters. As a result, the LLM often fails to initialize the correct objects and produces \harnesses that do not compile. After receiving feedback such as \textit{cannot find symbol} (often caused by hallucinated types or variables), the LLM attempts to fix the issue by trying alternative object constructions. However, these attempts still fail to satisfy the required object states, and the \harnesses remain non-compilable.

Path information is the most impactful prompt component: it prevents the LLM from generating brittle reflection-based \harnesses and constrains it to follow a statically feasible invocation path. \toolName's static analysis phase to extract this contextual information is therefore essential before attempting \harness generation.
Iterative targeted feedback is also effective and helps the LLM fix hallucinations.

\begin{tcolorbox}[boxrule=1pt,arc=.3em, left=2pt, right=2pt]
  \textbf{Answer to RQ3}:  \toolName’s design decisions all positively impact its effectiveness and lead to the generation of more successful \harnesses. Call-path details and entry-point context in the prompt, combined with shortest path selection and iterative feedback, guide the LLM to successfully produce many \harnesses across iterations.
\end{tcolorbox}

\subsection*{\textbf{RQ4: \rqsemgrep}}

\begin{table*}
\scriptsize
\centering
\captionsetup{font=small}
\caption{The number of unique CVEs that \semgrep classifies as reachable or undetermined in each module, along with their static reachability according to \toolName. \textsc{Strong reach. acc. \semgrep} and \textsc{Loose reach. acc. \semgrep} present the number of CVEs that \semgrep classifies as reachable with and without well-defined pattern matching, respectively. \textsc{Undet. acc. \semgrep} presents the number of CVEs that \semgrep classifies as having undetermined reachability.
\textsc{\symboltpl Call site present} reports the number of call sites associated with CVEs that exist in projects, regardless of their reachability. \textsc{Static. reach. acc. Fika} shows the number of call sites that \toolName determines to be statically reachable.}
\label{tab:rq5}

\begin{tabular}{|l||r|r|r||r|r|r||r|r|r|}
\hline

\textsc{\makecell[r]{Module}} &
\textsc{\makecell{Strong \\ reach. \\ acc. \\ SemGrep}} &
\textsc{\makecell{\symboltpl \\ Call site \\ present}} &
\textsc{\makecell{Static \\ reach. \\ acc. \\ Fika}} &
\textsc{\makecell{Loose \\ reach. \\ acc. \\ SemGrep}} &
\textsc{\makecell{\symboltpl \\ Call site \\ present}} &
\textsc{\makecell{Static \\ reach. \\ acc. \\ Fika}} &
\textsc{\makecell{Undet. \\ acc. \\ SemGrep}} &
\textsc{\makecell{\symboltpl \\ Call site \\ present}} &
\textsc{\makecell{Static \\ reach. \\ acc. \\ Fika}} \\
\hline

\rowcolor{rowGray}
\texttt{\href{https://github.com/apache/flink/tree/8af02592472e2a85eb3eeb81457ac3e6660d27e4/flink-formats/flink-parquet}{flink-parquet}} &
\nbSRsemgrep{2} &
\nbDir{2} &
\nbEfika{2} &
\nbLRsemgrep{5} &
\nbDir{0} &
\nbEfika{0} &
\nbUsemgrep{20} &
\nbDir{1} &
\nbEfika{1} \\

\texttt{\href{https://github.com/apache/flink/tree/8af02592472e2a85eb3eeb81457ac3e6660d27e4/flink-runtime}{flink-runtime}} &
\nbSRsemgrep{2} &
\nbDir{2} &
\nbEfika{2} &
\nbLRsemgrep{5} &
\nbDir{0} &
\nbEfika{0} &
\nbUsemgrep{24} &
\nbDir{2} &
\nbEfika{2} \\

\rowcolor{rowGray}
\texttt{\href{https://github.com/apache/flink/tree/8af02592472e2a85eb3eeb81457ac3e6660d27e4/flink-formats/flink-protobuf}{flink-protobuf}} &
\nbSRsemgrep{1} &
\nbDir{1} &
\nbEfika{1} &
\nbLRsemgrep{0} &
\nbDir{0} &
\nbEfika{0} &
\nbUsemgrep{5} &
\nbDir{0} &
\nbEfika{0} \\

\texttt{\href{https://github.com/apache/flink/tree/8af02592472e2a85eb3eeb81457ac3e6660d27e4/flink-formats/flink-orc}{flink-orc}} &
\nbSRsemgrep{1} &
\nbDir{1} &
\nbEfika{1} &
\nbLRsemgrep{4} &
\nbDir{1} &
\nbEfika{1} &
\nbUsemgrep{21} &
\nbDir{3} &
\nbEfika{2} \\

\rowcolor{rowGray}
\texttt{\href{https://github.com/apache/flink/tree/8af02592472e2a85eb3eeb81457ac3e6660d27e4/flink-metrics/flink-metrics-datadog}{flink-metrics-datadog}} &
\nbSRsemgrep{1} &
\nbDir{1} &
\nbEfika{1} &
\nbLRsemgrep{0} &
\nbDir{0} &
\nbEfika{0} &
\nbUsemgrep{3} &
\nbDir{0} &
\nbEfika{0} \\

\texttt{\href{https://github.com/graphhopper/graphhopper/tree/c721fa322f448976355951504525353d36bd6c36/core}{graphhopper-core}} &
\nbSRsemgrep{1} &
\nbDir{1} &
\nbEfika{1} &
\nbLRsemgrep{2} &
\nbDir{2} &
\nbEfika{2} &
\nbUsemgrep{5} &
\nbDir{1} &
\nbEfika{1} \\

\rowcolor{rowGray}
\texttt{\href{https://github.com/graphhopper/graphhopper/tree/c721fa322f448976355951504525353d36bd6c36/web-bundle}{graphhopper-web-bundle}} &
\nbSRsemgrep{1} &
\nbDir{1} &
\nbEfika{1} &
\nbLRsemgrep{2} &
\nbDir{2} &
\nbEfika{2} &
\nbUsemgrep{17} &
\nbDir{3} &
\nbEfika{3} \\

\texttt{\href{https://github.com/jooby-project/jooby/tree/7dc5b6ddaf1f2c251af04545c75a1e01cbfea899/modules/jooby-netty}{jooby-netty}} &
\nbSRsemgrep{2} &
\nbDir{2} &
\nbEfika{2} &
\nbLRsemgrep{0} &
\nbDir{0} &
\nbEfika{0} &
\nbUsemgrep{5} &
\nbDir{5} &
\nbEfika{5} \\

\rowcolor{rowGray}
\texttt{\href{https://github.com/jooby-project/jooby/tree/7dc5b6ddaf1f2c251af04545c75a1e01cbfea899/modules/jooby-jetty}{jooby-jetty}} &
\nbSRsemgrep{1} &
\nbDir{1} &
\nbEfika{1} &
\nbLRsemgrep{0} &
\nbDir{0} &
\nbEfika{0} &
\nbUsemgrep{6} &
\nbDir{6} &
\nbEfika{6} \\

\texttt{\href{https://github.com/jooby-project/jooby/tree/7dc5b6ddaf1f2c251af04545c75a1e01cbfea899/modules/jooby-http2-netty}{jooby-http2-netty}} & 
\nbSRsemgrep{1} &
\nbDir{1} &
\nbEfika{0} &
\nbLRsemgrep{0} &
\nbDir{0} &
\nbEfika{0} &
\nbUsemgrep{4} &
\nbDir{4} &
\nbEfika{1} \\

\rowcolor{rowGray}
\texttt{\href{https://github.com/jooby-project/jooby/tree/7dc5b6ddaf1f2c251af04545c75a1e01cbfea899/modules/jooby-http2-jetty}{jooby-http2-jetty}} &
\nbSRsemgrep{1} &
\nbDir{1} &
\nbEfika{1} &
\nbLRsemgrep{1} &
\nbDir{0} &
\nbEfika{0} &
\nbUsemgrep{7} &
\nbDir{4} &
\nbEfika{4} \\

\texttt{\href{https://github.com/jooby-project/jooby/tree/7dc5b6ddaf1f2c251af04545c75a1e01cbfea899/modules/jooby-utow}{jooby-utow}} &
\nbSRsemgrep{1} &
\nbDir{1} &
\nbEfika{1} &
\nbLRsemgrep{7} &
\nbDir{7} &
\nbEfika{7} &
\nbUsemgrep{3} &
\nbDir{3} &
\nbEfika{3} \\

\rowcolor{rowGray}
\rowcolor{rowGray}
\texttt{\href{https://github.com/Sayi/poi-tl/tree/58fdb6c7d9db7da53cd420ff1ebde49813bd7af2/poi-tl}{poi-tl}} & 
\nbSRsemgrep{1} &
\nbDir{1} &
\nbEfika{1} &
\nbLRsemgrep{0} &
\nbDir{0} &
\nbEfika{0} &
\nbUsemgrep{15} &
\nbDir{3} &
\nbEfika{3} \\

\hline
\texttt{Total} & 
\nbSRsemgrep{16} &
\nbDir{16} &
\nbEfika{15} &
\nbLRsemgrep{26} &
\nbDir{12} &
\nbEfika{12} &
\nbUsemgrep{135} &
\nbDir{35} &
\nbEfika{31} \\

\hline
\end{tabular}
\end{table*}

\begin{table*}
\scriptsize
\centering
\captionsetup{font=small}
\caption{Comparison of \toolName and \fuzzer for \executability analysis of statically reachable CVEs. A \newcheckmark indicates that the approach confirms the \executability of the vulnerable method within the given time budget. A \newcheckmark\textsubscript{900} indicates that \fuzzer covers the target only when its time budget is extended to 15 minutes (900\,s); a blank cell in the \fuzzer column indicates that \fuzzer fails even under that extended budget.}
\label{tab:rq5_42}

\newcounter{rowcount}

\begin{tabular}{|l|l|l|l|c|c|r|}
\hline

\textsc{\#} &
\textsc{CVE} &
\textsc{Module} &
\textsc{Vul TPL} &
\textsc{\makecell{Jazzer \\ (\toolName's \\ static anal. \\ + fuzzing)}} &
\textsc{\makecell{\toolName \\ (static anal. \\ + scenario \\ generation)}} &
\textsc{\makecell{Time \\ budget \\ (Sec.)}} \\

\hline

\rowcolor{rowGray}
\stepcounter{rowcount}\therowcount &
\texttt{\cve{CVE-2021-0341}} &
\texttt{\fmodule{flink-metrics-datadog}} &
\texttt{\ftpl{com.squareup.okhttp3:okhttp}} &
 & 
 &
160 / 900\\

\stepcounter{rowcount}\therowcount &
\texttt{\cve{CVE-2024-7254}} &
\texttt{\fmodule{flink-orc}} &
\texttt{\ftpl{com.google.protobuf:protobuf-java}} &
\newcheckmark &
\newcheckmark &
22 \\

\rowcolor{rowGray}
\stepcounter{rowcount}\therowcount &
\texttt{\cve{CVE-2021-22569}} &
\texttt{\fmodule{flink-orc}} &
\texttt{\ftpl{com.google.protobuf:protobuf-java}} &
\newcheckmark &
\newcheckmark &
22 \\

%%%%%%%%%%%%%%%%%%%%%%%%%%%%%
\stepcounter{rowcount}\therowcount &
\texttt{\cve{CVE-2024-23454}} &
\texttt{\fmodule{flink-orc}} &
\texttt{\ftpl{org.apache.hadoop:hadoop-common}} &
\newcheckmark &
\newcheckmark &
27 \\

\rowcolor{rowGray}
\stepcounter{rowcount}\therowcount &
\texttt{\cve{CVE-2022-3171}} &
\texttt{\fmodule{flink-orc}} &
\texttt{\ftpl{com.google.protobuf:protobuf-java}} &
\newcheckmark &
\newcheckmark &
22 \\

\stepcounter{rowcount}\therowcount &
\texttt{\cve{CVE-2024-23454}} &
\texttt{\fmodule{flink-parquet}} &
\texttt{\ftpl{org.apache.hadoop:hadoop-common}} &
\newcheckmark &
\newcheckmark &
140 \\

\rowcolor{rowGray}
\stepcounter{rowcount}\therowcount &
\texttt{\cve{CVE-2024-7254}} &
\texttt{\fmodule{flink-protobuf}} &
\texttt{\ftpl{com.google.protobuf:protobuf-java}} &
\newcheckmark &
\newcheckmark &
54 \\

\stepcounter{rowcount}\therowcount &
\texttt{\cve{CVE-2024-23454}} &
\texttt{\fmodule{flink-runtime}} &
\texttt{\ftpl{org.apache.hadoop:hadoop-common}} &
 & % jazzer not covered
\boldcheckmark &
27 / 900 \\

\rowcolor{rowGray}
\stepcounter{rowcount}\therowcount &
\texttt{\cve{CVE-2025-48924}} &
\texttt{\fmodule{flink-runtime}} &
\texttt{\ftpl{org.apache.commons:commons-lang3}} &
\newcheckmark &
\newcheckmark &
85 \\

\stepcounter{rowcount}\therowcount &
\texttt{\cve{CVE-2020-8908}} &
\texttt{\fmodule{graphhopper-web-bundle}} &
\texttt{\ftpl{com.google.guava:guava}} &
\newcheckmark\textsubscript{900} & % jazzer not covered
\boldcheckmark &
17 / 900 \\

\rowcolor{rowGray}
\stepcounter{rowcount}\therowcount &
\texttt{\cve{CVE-2023-2976}} &
\texttt{\fmodule{graphhopper-web-bundle}} &
\texttt{\ftpl{com.google.guava:guava}} &
\newcheckmark\textsubscript{900} & % jazzer not covered
\boldcheckmark &
17 / 900 \\

\stepcounter{rowcount}\therowcount &
\texttt{\cve{CVE-2025-5115}} &
\texttt{\fmodule{jooby-http2-jetty}} &
\texttt{\ftpl{org.eclipse.jetty.http2:http2-common}} &
\newcheckmark &
\newcheckmark &
18 \\

\rowcolor{rowGray}
\stepcounter{rowcount}\therowcount &
\texttt{\cve{CVE-2024-8184}} &
\texttt{\fmodule{jooby-http2-jetty}} &
\texttt{\ftpl{org.eclipse.jetty:jetty-server}} &
\newcheckmark &
\newcheckmark &
18 \\

\stepcounter{rowcount}\therowcount &
\texttt{\cve{CVE-2023-26049}} &
\texttt{\fmodule{jooby-http2-jetty}} &
\texttt{\ftpl{org.eclipse.jetty:jetty-server}} &
\newcheckmark &
\newcheckmark &
18 \\

\rowcolor{rowGray}
\stepcounter{rowcount}\therowcount &
\texttt{\cve{CVE-2023-26048}} &
\texttt{\fmodule{jooby-http2-jetty}} &
\texttt{\ftpl{org.eclipse.jetty:jetty-server}} &
\newcheckmark &
\newcheckmark &
18 \\

\stepcounter{rowcount}\therowcount &
\texttt{\cve{CVE-2023-44487}} &
\texttt{\fmodule{jooby-http2-jetty}} &
\texttt{\ftpl{org.eclipse.jetty.http2:http2-server}} &
\newcheckmark &
\newcheckmark &
18 \\

\rowcolor{rowGray}
\stepcounter{rowcount}\therowcount &
\texttt{\cve{CVE-2023-34462}} &
\texttt{\fmodule{jooby-http2-netty}} &
\texttt{\ftpl{io.netty:netty-handler}} &
 & % jazzer not covered
\boldcheckmark &
141 / 900 \\

\stepcounter{rowcount}\therowcount &
\texttt{\cve{CVE-2024-13009}} &
\texttt{jooby-jetty} &
\texttt{\ftpl{org.eclipse.jetty:jetty-server}} &
 & % jazzer not covered
\boldcheckmark &
81 / 900 \\

\rowcolor{rowGray}
\stepcounter{rowcount}\therowcount &
\texttt{\cve{CVE-2024-8184}} &
\texttt{\fmodule{jooby-jetty}} &
\texttt{\ftpl{org.eclipse.jetty:jetty-server}} &
\newcheckmark &
\newcheckmark &
33 \\

\stepcounter{rowcount}\therowcount &
\texttt{\cve{CVE-2023-26049}} &
\texttt{\fmodule{jooby-jetty}} &
\texttt{\ftpl{org.eclipse.jetty:jetty-server}} &
\newcheckmark &
\newcheckmark &
33 \\

\rowcolor{rowGray}
\stepcounter{rowcount}\therowcount &
\texttt{\cve{CVE-2023-26048}} &
\texttt{\fmodule{jooby-jetty}} &
\texttt{\ftpl{org.eclipse.jetty:jetty-server}} &
\newcheckmark &
\newcheckmark &
33 \\

\stepcounter{rowcount}\therowcount &
\texttt{\cve{CVE-2023-40167}} &
\texttt{\fmodule{jooby-jetty}} &
\texttt{\ftpl{org.eclipse.jetty:jetty-http}} &
\newcheckmark &
\newcheckmark &
141 \\

\rowcolor{rowGray}
\stepcounter{rowcount}\therowcount &
\texttt{\cve{CVE-2024-6763}} &
\texttt{\fmodule{jooby-jetty}} &
\texttt{\ftpl{org.eclipse.jetty:jetty-http}} &
\newcheckmark &
\newcheckmark &
141 \\

\stepcounter{rowcount}\therowcount &
\texttt{\cve{CVE-2025-11143}} &
\texttt{\fmodule{jooby-jetty}} &
\texttt{\ftpl{org.eclipse.jetty:jetty-http}} &
\newcheckmark &
\newcheckmark &
141 \\

\rowcolor{rowGray}
\stepcounter{rowcount}\therowcount &
\texttt{\cve{CVE-2025-58056}} &
\texttt{jooby-netty} &
\texttt{\ftpl{io.netty:netty-codec-http}} &
 & % jazzer not covered
\boldcheckmark &
150 / 900 \\

\stepcounter{rowcount}\therowcount &
\texttt{\cve{CVE-2024-29025}} &
\texttt{\fmodule{jooby-netty}} &
\texttt{\ftpl{io.netty:netty-codec-http}} &
\newcheckmark &
\newcheckmark &
59 \\

\rowcolor{rowGray}
\stepcounter{rowcount}\therowcount &
\texttt{\cve{CVE-2025-67735}} &
\texttt{\fmodule{jooby-netty}} &
\texttt{\ftpl{io.netty:netty-codec-http}} &
\newcheckmark &
\newcheckmark &
59 \\

\stepcounter{rowcount}\therowcount &
\texttt{\cve{CVE-2025-58057}} &
\texttt{\fmodule{jooby-netty}} &
\texttt{\ftpl{io.netty:netty-codec}} &
\newcheckmark &
\newcheckmark &
118\\

\rowcolor{rowGray}
\stepcounter{rowcount}\therowcount &
\texttt{\cve{CVE-2023-34462}} &
\texttt{\fmodule{jooby-netty}} &
\texttt{\ftpl{io.netty:netty-handler}} &
\newcheckmark &
\newcheckmark &
132\\

\stepcounter{rowcount}\therowcount &
\texttt{\cve{CVE-2025-25193}} &
\texttt{\fmodule{jooby-netty}} &
\texttt{\ftpl{io.netty:netty-common}} &
\newcheckmark &
\newcheckmark &
148\\

\rowcolor{rowGray}
\stepcounter{rowcount}\therowcount &
\texttt{\cve{CVE-2023-4639}} &
\texttt{jooby-utow} &
\texttt{\ftpl{io.undertow:undertow-core}} &
\newcheckmark &
\newcheckmark &
126\\

%without pattern semgrep itself says reachable
\stepcounter{rowcount}\therowcount &
\texttt{\cve{CVE-2024-1635}} &
\texttt{jooby-utow} &
\texttt{\ftpl{io.undertow:undertow-core}} &
\newcheckmark &
\newcheckmark &
126 \\

%without pattern semgrep itself says reachable
\rowcolor{rowGray}
\stepcounter{rowcount}\therowcount &
\texttt{\cve{CVE-2024-5971}} &
\texttt{jooby-utow} &
\texttt{\ftpl{io.undertow:undertow-core}} &
\newcheckmark &
\newcheckmark &
126 \\

%without pattern semgrep itself says reachable
\stepcounter{rowcount}\therowcount &
\texttt{\cve{CVE-2025-12543}} &
\texttt{jooby-utow} &
\texttt{\ftpl{io.undertow:undertow-core}} &
\newcheckmark &
\newcheckmark &
126 \\

%without pattern semgrep itself says reachable
\rowcolor{rowGray}
\stepcounter{rowcount}\therowcount &
\texttt{\cve{CVE-2023-1108}} &
\texttt{jooby-utow} &
\texttt{\ftpl{io.undertow:undertow-core}} &
\newcheckmark &
\newcheckmark &
126 \\

%without pattern semgrep itself says reachable
\stepcounter{rowcount}\therowcount &
\texttt{\cve{CVE-2024-4027}} &
\texttt{jooby-utow} &
\texttt{\ftpl{io.undertow:undertow-core}} &
\newcheckmark &
\newcheckmark &
126 \\

%without pattern semgrep itself says reachable
\rowcolor{rowGray}
\stepcounter{rowcount}\therowcount &
\texttt{\cve{CVE-2024-3884}} &
\texttt{jooby-utow} &
\texttt{\ftpl{io.undertow:undertow-core}} &
\newcheckmark &
\newcheckmark &
126 \\

%without pattern semgrep itself says reachable
\stepcounter{rowcount}\therowcount &
\texttt{\cve{CVE-2024-6162}} &
\texttt{jooby-utow} &
\texttt{\ftpl{io.undertow:undertow-core}} &
\newcheckmark &
\newcheckmark &
126 \\

\rowcolor{rowGray}
\stepcounter{rowcount}\therowcount &
\texttt{\cve{CVE-2023-1973}} &
\texttt{\fmodule{jooby-utow}} &
\texttt{\ftpl{io.undertow:undertow-core}} &
\newcheckmark &
\newcheckmark &
126\\

\stepcounter{rowcount}\therowcount &
\texttt{\cve{CVE-2024-3653}} &
\texttt{\fmodule{jooby-utow}} &
\texttt{\ftpl{io.undertow:undertow-core}} &
\newcheckmark &
\newcheckmark &
126\\

\rowcolor{rowGray}
\stepcounter{rowcount}\therowcount &
\texttt{\cve{CVE-2024-1459}} &
\texttt{\fmodule{jooby-utow}} &
\texttt{\ftpl{io.undertow:undertow-core}} &
\newcheckmark &
\newcheckmark &
126\\

\stepcounter{rowcount}\therowcount &
\texttt{\cve{CVE-2025-31672}} &
\texttt{\fmodule{poi-tl}} &
\texttt{\ftpl{org.apache.poi:poi-ooxml}} &
\newcheckmark &
\newcheckmark &
30\\

\rowcolor{rowGray}
\stepcounter{rowcount}\therowcount &
\texttt{\cve{CVE-2025-48924}} &
\texttt{\fmodule{poi-tl}} &
\texttt{\ftpl{org.apache.commons:commons-lang3}} &
\newcheckmark &
\newcheckmark &
58\\

\stepcounter{rowcount}\therowcount &
\texttt{\cve{CVE-2024-38808}} &
\texttt{\fmodule{poi-tl}} &
\texttt{\ftpl{org.springframework:spring-expression}} &
 & % jazzer not covered
\boldcheckmark &
22 / 900\\

\hline
\multicolumn{4}{|r|}{\textsc{Total}} &
36 (38) &
43 &
1hr / 3hr\\

\hline
\end{tabular}
\end{table*}

With this research question, we evaluate the usefulness of \toolName for vulnerability reachability analysis and compare it with state of the art reachability analysis tools.
In \textit{RQ4.a}, we discuss the results from the \toolName's static analysis comparison with the pattern based reachability analysis of \semgrep. In \textit{RQ4.b}, we discuss the results from the \toolName's dynamic analysis comparison with \fuzzer.

\paragraph*{RQ4.a: Comparison of \toolName's static analysis with pattern based reachability analysis}

In \autoref{tab:rq5}, we report the results of \semgrep's CVE reachability analysis as follows.
In column 2, \textsc{Strong reach. acc. Semgrep} refers to \semgrep matching a specific code pattern to determine reachability of a vulnerable library method; in column 5, \textsc{Loose reach. acc. Semgrep}, refers to reachability determined by the presence of the \dependency in the dependency tree according to \semgrep; in column 8, \textsc{Undet. Acc. \semgrep} refers to \semgrep reporting the reachability as undetermined.
For each \semgrep diagnosis, we also collect two metrics from \toolName: \textsc{\symboltpl call site present}, the number of CVEs for which \toolName can statically determine at least one call site; and \textsc{Static. Reach. Acc. \toolName}, the number of CVEs for which \toolName can find a static call path.

\semgrep identifies 16 CVEs as reachable based on a well-defined pattern of vulnerable \symboltpl usage, which is strong static reachability evidence (modulo dead code).
For 15 CVEs, \toolName statically identifies the call site containing the corresponding \symboltpl and augments this with statically reachable path information, and does not find a path to reach 1 CVE.

\semgrep's loose reachability analysis reports 26 reachable CVEs across 7 modules, based on the presence of the \dependency in the dependency tree.
By opposition to this loose diagnosis, \toolName reports static reachability evidence for 12 of these CVEs across 4 modules.
When \semgrep leaves developers uncertain about whether a vulnerable dependency is actually used, \toolName pinpoints the call sites and provides static reachable paths for them for further dynamic analysis.

\semgrep classifies 135 CVEs as undetermined across all 13 modules.
For these CVEs, \semgrep rules do not contain patterns to determine reachability, and developers cannot act upon this diagnosis without significant additional manual analysis.
Out of the same 135 CVEs, \toolName finds that 35 are invoked in the project, and finds static paths to reach 31 CVEs.
Altogether, \toolName finds 58 CVEs as reachable through static analysis.

\modified{
\paragraph*{RQ4.b: Comparison of \toolName's dynamic analysis with fuzzing}
For the 58 CVEs for which \toolName finds statically reachable paths, the corresponding project test suites already provide \executability guarantees for 14. For the remaining 44 CVEs, we run both the \testGenEngine of \toolName and \fuzzer to determine whether the vulnerable methods can actually be executed. We provide the data extracted through the static analysis phase of \toolName to both \fuzzer and \testGenEngine, to direct them towards maximizing coverage for the target \dependency call site.

In \autoref{tab:rq5_42}, we present the \executability results obtained through \toolName and \fuzzer. The \newcheckmark indicates whether the \executability of the vulnerable \symboltpl is detected by \toolName and \fuzzer. The last column reports the time taken by \toolName for the \harness generation, which is also the time allocated to \fuzzer.

For \projFlink, \toolName provides \executability guarantees for 7 of 8 CVEs across 5 modules. \fuzzer provides guarantees for 6 of them under the same time budget, but fails to find the \executability of \texttt{CVE-2024-23454} in the \dependency\xspace \texttt{hadoop-common} (row 8).
Neither \toolName nor \fuzzer provides guarantees of \executability for the vulnerable method related to \texttt{CVE-2021-0341} (row 1), as it requires complex object initialization. For \texttt{graphhopper-web-bundle} (rows 10 and 11), \toolName generates successful \harnesses in 17 seconds. Within this time budget \fuzzer fails to generate inputs demonstrating the \executability of \texttt{CVE-2020-8908} and \texttt{CVE-2023-2976}. In projects \texttt{jooby} and \texttt{poi-tl}, \toolName provides guarantees of \executability for 30 and 3 CVEs respectively, but \fuzzer only provides guarantees for 27 and 2.

Across the full dynamic evaluation, \toolName confirms the \executability of 43 out of 44 CVEs (98\%), compared with 36 out of 44 (82\%) for \fuzzer under the same time budget. To check whether \fuzzer simply needs more time to reach the 8 CVEs it missed, we re-run it on each of them with an extended budget of 15 minutes. Under this budget, \fuzzer covers two more CVEs in \texttt{graphhopper-web-bundle} (rows 10 and 11, marked \newcheckmark\textsubscript{900} in the table), but still fails on the remaining 6. 
%It fails to generate valid arguments for the methods along the path to the target in \texttt{jooby-http2-netty} (row 17) and \texttt{flink-runtime} (row 8), and similarly fails for the CVEs in \texttt{flink-metrics-datadog} (row 1), \texttt{jooby-jetty} (row 18), \texttt{jooby-netty} (row 25), and \texttt{poi-tl} (row 44).
Even with this extended budget, \toolName retains an 11\% coverage advantage over \fuzzer (43/44 against 38/44), showing that it is substantially more effective at establishing the \executability of vulnerable methods, both in time budget and in coverage.

Beyond coverage, \toolName further produces a directly executable and human-readable \harness, whereas \fuzzer produces inputs only as part of its fuzzing corpus. \toolName therefore provides both higher reachability coverage and more reproducible evidence: its output can be integrated into the build pipeline with no additional dependency, but the inputs generated by \fuzzer can only be executed through the \fuzzer engine.
}

\begin{tcolorbox}[boxrule=1pt,arc=.3em, left=2pt, right=2pt]
\textbf{Answer to RQ4}:
\toolName is the state-of-the-art for providing \executability guarantees of CVEs, clearly improving over pattern-based reachability analysis with \semgrep\xspace \modified{and coverage-guided fuzzing with \fuzzer}. These \executability guarantees, demonstrated with \executable \harnesses, let developers make prioritization decisions regarding the remediation of their dependency attack surface.
\end{tcolorbox}

\section{Threats to validity}%

One main threat to validity of our study is the dependence on static analysis to identify the paths to \dependency method calls. Some invocations of \dependencies may occur only at runtime or may be obfuscated, meaning our static analysis could miss them. This threat is minimized by our dynamic analysis step, where we invoke each statically identified method dynamically, increasing the chances of indirectly reaching other statically invisible functionality. 

Our choice of the prompt and the number of iterations of the feedback loop are another threat to validity of the \harness generation pipeline. To mitigate this, we evaluate the impact of our design decisions through an incremental ablation study. 
The choice of the LLM also impacts the results, and a more powerful LLM could result in better outcomes. However, as the main goal of implementing \toolName is not to evaluate the ability of different LLMs to generate \harnesses, but rather to provide a proof of concept, we do not experiment with different LLMs. 
% With a larger context window and cost budget, it is possible to explore concatenating feedback from all iterations rather than only the most recent one.

The final threat to validity concerns the transferability of our approach to other Java ecosystems such as Kotlin and to other Java testing frameworks such as TestNG. In the latter case, our approach requires only minimal changes, specifically adapting the prompt to the LLM. In the former case, our approach or findings may not be directly transferable. \modified{Extending \toolName to other languages such as Python or JavaScript would require reimplementing the static analysis, the prompt construction, and the \harness validation, as these components rely on Java-specific tools such as SootUp, Spoon, and JaCoCo.}

\section{Related work}

To the best of our knowledge, no prior tool generates \harnesses specifically for \executability analysis of \dependencies. In this section, we discuss previous work relevant to our approach, including studies on program reachability analysis and dynamic analysis of \dependencies.

A few recent works on program reachability analysis focus on improving the scalability and efficiency of call graph generation in the presence of \dependencies. Keshani \etal \cite{keshani2024frankenstein} introduce a fast and lightweight approach called Frankenstein for complete static call graph generation, which processes \dependencies individually and stitches them together afterward. Similarly, Wang \etal \cite{wang2025reachcheck} propose a library-aware reachability analysis tool that is efficient enough to be integrated into IDEs.
Although such static reachability analysis methods are efficient, they suffer from imprecision and unsoundness \cite{samhi2024dynamic}, \cite{samhi2025runtimecall}. 
\toolName performs dynamic analysis guided by static analysis and LLMs to overcome the limitations of purely static reachability analysis.

\modified{
Symbolic execution and formal verification can also target program reachability. Previous work has applied these techniques to determine whether input controlled by an advanced attacker can drive program execution to a specific vulnerable location \cite{ducousso2023adversarial}, construct sound and complete bug finders \cite{correnson2023bugfinder}, and quantify difficulty of triggering bugs \cite{girol2024quantifybug}. These approaches provide formal guarantees, but do not scale to the full dependency graphs of real-world projects. \toolName provides a practical solution to synthesize executable code that demonstrates reachability empirically.
}

\modified{
Directed greybox fuzzing is another well-established approach for reaching specific code locations. AFLGo \cite{bohme2017aflgo} proposes the idea of guiding fuzzing toward target sites by using distance metrics over the call graph. Building on this approach, BEACON \cite{huang2022beacon} uses static analysis to prune infeasible paths before fuzzing to reduce redundant effort, SelectFuzz \cite{lyu2023selectfuzz} selectively instruments only relevant paths to improve scalability, and BinGo \cite{lin2025deviate} uses probabilistic reachability analysis to guide the fuzzer when control flow deviates from the expected path. Along the same lines, Sapia \etal \cite{sapia2026reachability} identify the challenge of creating both the required initial environment and generating inputs that can reach internal functions, and propose a pipeline that combines static analysis and LLM agents with directed fuzzing to address this at scale. In the Java ecosystem specifically, \fuzzer \cite{dechand2026jazzer} adapts coverage-guided fuzzing for the JVM through bytecode instrumentation and Java-specific vulnerability detectors. These fuzzers generate raw inputs that maximize coverage toward a target, whereas \toolName generates structured, human-readable code that satisfies the complex object initialization often required by Java methods. We compare \toolName against \fuzzer in RQ4 and show that \toolName reaches more vulnerable \dependency call sites.
}

Hybrid approaches that combine both static and dynamic reachability analysis are used by a few other related tools.
Mockingbird \cite{holland2019mockingbird}, a tool to analyze Java programs, uses static analysis to identify potentially vulnerable code regions before dynamically executing those regions using mock objects. 
GAPS \cite{doria2026mindgapsbridginggaps} analyzes method reachability in Android applications through a hybrid framework that uses static call graph traversal to guide dynamic execution. 
\toolName also aims to benefit from both static and dynamic reachability analysis. 
\toolName specifically targets the \dependency call sites, and aims to provide strong guarantees of \executability for dependency usage. 

When it comes to the vulnerability analysis of \dependencies, ecosystem-scale studies show that even though many projects depend on vulnerable libraries, only a very small fraction actually include call paths that reach the vulnerable functions \cite{mir2023reachability}. 
Grgic \etal \cite{grgic2026propagation} evaluate four open-source SBOM tools and find that, while they identify vulnerable components, they lack support for code reachability analysis.
Ponta \etal \cite{ponta2020detection} propose Steady, a tool that determines if the vulnerable parts of a \dependency are actually reachable. It monitors code through test execution to see which parts are used.
% However, its effectiveness depends on the completeness of the existing test suites. 
In our work, we find that even well-tested client projects cover, on average, only 55\% of \dependency call sites. This limited coverage reduces the tool’s ability to detect dynamically reachable vulnerabilities.

Vulnerability exploitation tools also use dynamic analysis to target and exercise vulnerable functions.
Transfer, a tool introduced by Kang \etal \cite{kang2022transfer}, uses test mimicry to create vulnerability exploits by copying specific program states and inputs from a library’s own tests to check whether a client project can reach those same states. Vesta \cite{chen2024vultestlib} improves on this by extracting parameters directly from known exploit code and injecting those values into a client project’s tests to confirm exploitability. Magneto \unskip ~\cite{chou2024fuzzingreachability} further handles complex scenarios where a vulnerable function is buried deep within long call chains. It uses a step-wise approach that breaks the chain into individual hops and processes each hop separately. Zhao \etal \cite{zhao2026triggeringdetectingexploitablelibrary} propose a similar fuzzing-based approach for exploiting vulnerabilities in C/C++.
Compared to these tools, \toolName focuses on analyzing the \executability of \dependency methods rather than exploiting vulnerabilities. Instead of relying on existing exploit code or test artifacts, we extract execution context directly from the client project. We synthesize complete, executable code that triggers the full call chain end-to-end, rather than exploring it incrementally. This allows \toolName to provide guarantees of \executability for dependency call sites.

Apart from vulnerability analysis, \dependency reachability is often performed indirectly in broader dependency analysis work. Studies that perform dynamic analysis of \dependencies typically identify the usage of \dependencies as a preliminary step, through the execution of existing tests.
Soto-Valero \etal \cite{cesar2023coverage} propose a methodology for removing unused \dependencies based on test coverage reports. However, as many projects do not have comprehensive test suites, this approach can lead to incorrectly removing dependencies whose methods are not executed during testing but are still required in production, potentially leading to failures.
Amusuo \etal \cite{amuso2025ztd} use test execution to trigger runtime \dependency interactions and determine the resource accesses exercised by \dependencies. In this case as well, weak test suites can result in incomplete results.
% To improve the coverage of \dependency methods, Chen \etal \cite{chen2020testing}, and Vikram \etal \cite{vikram2025freshnesspinneddependenciesmaven} suggest merging the test suites of downstream applications that use a specific \dependency method. Their approaches, though, cannot guarantee method coverage for unpopular \dependencies.
Jayasuriya \etal \cite{jayasuriya2024breaking} study the behavioral breaking changes in \dependencies by running the client project test suites. The authors mention that about 40\% of all projects do not have a test suite, and even those that do rarely cover all \dependency call sites. From another perspective, Raj \etal \cite{raj2025supportingopensourcelibrary} find that only 32\% of client projects use \dependency methods that are covered by library test suites. 
% This represents a significant limitation for empirical studies that rely on client project test suites to observe \dependency interactions.
With \toolName, we aim to address these limitations and improve dynamic analysis of \dependency behavior by generating \harnesses that can cover a wider range of \dependency methods.

\section{Conclusion}

In this paper, we propose the tool \toolName to identify \executable \dependency call sites in a project and provide guarantees of their \executability. To determine \executability, \toolName runs developer-written test suites and identifies the call sites that are already proven to be \executable. For the call sites that are not covered by the existing tests, \toolName extracts the paths required to reach them, along with path-related context, through static analysis. Then, using this rich context, \toolName prompts an LLM to generate a \harness that can trigger the non-covered call sites and serve as concrete evidence of their \executability.
To evaluate our approach, we select eight real-world Java Maven projects, analyze their \dependency usage, and generate \harnesses for the non-covered \dependency call sites. Our results indicate that around half of the \dependency call sites are not covered even by well-written developer test suites. With \toolName, we show that, when provided with rich context, LLMs can  cover previously non-covered call sites, even those deep in the codebase. Overall, \toolName increases \executability guarantees of \dependency call sites by \improvement.

We also show that \toolName can improve state-of-the-art vulnerability scanning tools such as \semgrep. \modified{\toolName goes beyond static approximations of \semgrep by exposing real execution paths to vulnerable functions and improves over \fuzzer, a coverage-guided directed fuzzer. \toolName confirms the \executability of 98\% of targeted vulnerabilities, compared to 82\% for \fuzzer under the same time budget. \toolName reduces uncertainty and provides human-readable, actionable evidence of vulnerability reachability.}
Building on this work, it is possible to improve the accuracy of dependency analysis tools across different use cases. We identify four such applications: revealing runtime access privileges of \dependencies (such as file or network access), detecting breaking library updates, improving the accuracy of coverage-based debloating or \dependency specialization tools, and identifying implicit \dependencies. %In future work, we plan to explore applying \toolName to other applications.

%In our paper, we generate unit tests with LLMs that specifically target \dependencies.

\section{Data Availability Statement} 

The implementation of \toolName\footnote{\href{https://visitsweden.com/what-to-do/food-drink/swedish-kitchen/all-about-swedish-fika/}{Fika} – the social Swedish coffee break – is a cherished tradition and a cornerstone of daily life in Sweden} is publicly available on \href{https://github.com/sparkrew/fika}{GitHub}. The evaluation data, including project configurations, and execution results is available on \href{https://github.com/sparkrew/fika/blob/main/Experiments.md}{GitHub}. The results of the comparison with \semgrep and \fuzzer are also available on \href{https://github.com/sparkrew/fika/blob/main/BaselineExperiments.md}{GitHub}.

\section{Acknowledgment} 
We thank Shin Hwei Tan for feedback on the early versions of this paper. 
This work is supported by the CHAINS project funded by Swedish Foundation for Strategic Research (SSF), Natural Sciences and Engineering Research Council of Canada (NSERC), Fonds de Recherche du Québec (FRQ) and the Canada Research Chairs Program.

\balance
\bibliographystyle{ieeetr}
\bibliography{main}

\end{document}